\documentclass[11pt,
    letterpaper,
    reprint,
    notitlepage,
    superscriptaddress,
    aps,pra]{revtex4-2}
\usepackage{notes2bib}
\usepackage{natbib}
\usepackage{gensymb}
\usepackage{amsthm}
\usepackage{amsmath} 
\usepackage{color}
\usepackage{bm}
\usepackage{amsmath,amssymb}
\usepackage{amsfonts}
\usepackage{dsfont}
\usepackage{float}
\usepackage{graphicx}
\usepackage{float}
\usepackage{subfig}
\usepackage{verbatim}
\usepackage{amsfonts}
\usepackage{bbold}
\usepackage{dcolumn}
\newcommand{\h}{\hat}
\usepackage{bm}
\usepackage{soul}
\usepackage[dvipsnames]{xcolor}
\usepackage{listings}
\definecolor{blueprl}{RGB}{46,48,146}
\usepackage[pdftex,colorlinks=true,urlcolor=blueprl,citecolor=blueprl,linkcolor=blueprl]{hyperref}
\usepackage{times}
\usepackage{color}
\usepackage[dvipsnames]{xcolor}

\usepackage{physics}
\usepackage{caption}
\usepackage{subcaption}

\usepackage[mathscr]{euscript}
\usepackage{amsmath}
\usepackage{graphicx}
\usepackage{dcolumn}
\usepackage{bm}
\usepackage{epsfig}
\usepackage{amssymb,latexsym,mathrsfs}
\usepackage{graphicx}
\usepackage{color}
\usepackage{hyperref}
\usepackage{float}
\usepackage{diagbox}
\usepackage{ulem}

\newcommand{\vt}{\vphantom{\frac{1}{2}}}

\newcommand{\p}{\partial}
\newcommand{\mrm}[1]{\mathrm{#1}}
\usepackage{amsmath,mathtools}

\begin{document}
 \title{Quantum metasurfaces as probes of vacuum particle content}
 \author{Germain Tobar}
 \email{germain.tobar@fysik.su.se}
\affiliation{Department of Physics, Stockholm University, SE-106 91 Stockholm, Sweden}

\author{Joshua Foo}
\affiliation{Department of Physics and Astronomy, University of Waterloo, Waterloo, Ontario, Canada, N2L 3G1}

\author{Sofia Qvarfort}
\affiliation{Department of Physics, Stockholm University, SE-106 91 Stockholm, Sweden}
\affiliation{Nordita, KTH Royal Institute of Technology and Stockholm University,
Hannes Alfvens vag 12, SE-114 19 Stockholm, Sweden}

\author{Fabio Costa}
\affiliation{Nordita, KTH Royal Institute of Technology and Stockholm University,
Hannes Alfvens vag 12, SE-114 19 Stockholm, Sweden}

\author{Rivka Bekenstein}
\affiliation{Racah Institute of Physics, The Hebrew University of Jerusalem, Jerusalem 91904, Israel}

\author{Magdalena Zych}
\email{magdalena.zych@fysik.su.se}
\affiliation{Department of Physics, Stockholm University, SE-106 91 Stockholm, Sweden}

\begin{abstract} 
The quantum vacuum of the electromagnetic field is inherently entangled across distinct spatial sub-regions, resulting in entangled particle content across these sub-regions. However, accessing this particle content in a controlled laboratory experiment has remained out of experimental reach. Here, we analyse the feasibility of witnessing such vacuum effects with highly non-perturbative boundary condition changes with a quantum metasurface made from a two-dimensional sub-wavelength atomic array. The array's response to light is tunable between transmissive and reflective states by a control atom that is excited to a Rydberg state. We find that vacuum photon content from non-perturbative changes of the boundary conditions and therefore distinct spatial sub-regions of the vacuum causes subtle frequency shifts and analyse the accessibility to sub-wavelength atom array platforms. This novel approach to probing vacuum particle content leverages the system’s unique ability to generate coherent dynamics in superpositions of transmissive and reflective states of the array's reflectivity, providing a quantum-enhanced platform for observing vacuum particle creation driven by highly non-perturbative changes in the boundary conditions of the electromagnetic vacuum. 
\end{abstract} 

\date{\today} 

\maketitle 

\section{Introduction}
In any relativistic quantum field theory, including quantum electrodynamics (QED), local regions of the field are not in the vacuum state even when the global field is in the vacuum \cite{ReehH.1961Bzuv,SummersStephenJ.1985TvvB,RevModPhys.90.045003,Casini_2009,VAZQUEZ2014112}. This is because vacuum states of field theories contain local fluctuations. The key physical consequence of this phenomenon is the prediction of non-zero particle content in sub-regions of a quantum field that is in its global vacuum state. Various cavity QED experiments have explored the unique properties of the QED vacuum, such as the cooperative Lamb shift observed in atomic ensembles \cite{KeaveneyJ2012CLsi,HutsonRossB2024Oomc}. A significant challenge in modern vacuum QED experiments is detecting particle content from the vacuum, where photons from spatial sub-regions of the vacuum become observable. This phenomenon occurs in a QED system when for example a mirror’s motion alters the electromagnetic field vacuum mode profiles, known as the dynamical Casimir effect (DCE) \cite{MooreGeraldT.1970QTot, FullingS.A.1976RfaM, DaviesP.C.W.1977RfMM, Birrell_Davies_1982}. When the mirror moves at sufficiently high speeds, vacuum modes fail to adiabatically adjust to the mirror’s new position, leading to a mismatch between the updated cavity modes and the original quantum state. This mismatch generates particle content, which, if detected, would correspond to photons spontaneously produced from highly non-perturbative spatial boundary condition changes.

The experimental observation of this phenomenon is a key objective for experiments at the interface of relativity and quantum theory because particle creation from the vacuum is also closely related to several fundamental yet unobserved phenomena such as Unruh and Hawking radiation \cite{HAWKINGS.W1974Bhe,WadiaSpentaR.2021PCbB,PhysRevD.14.870,RevModPhys.80.787}, which have recently received more attention for experimental relevance \cite{ZhengHong-Tao2025EaUe,deswal2025timeresolvedsuperradiantlyamplifiedunruh}. Such phenomena bridge several disciplines including quantum optics \cite{Su_2017, alma991015317318203131} and quantum field theory in curved spacetime \cite{dodonovphysics2010007,CHUNG1994305}, with applications to analogue gravity \cite{holzhey1994geometric, goodPhysRevD.101.025012,Akal_2021}. In this way, unambiguous confirmation of the local particle content due to highly non-perturbative boundary condition changes, would mark a major milestone in experiments in relativistic quantum physics.

The practicality of observing vacuum particle creation with a physical mirror has been debated, since a macroscopic object moving at the required relativistic speeds would endure immense mechanical stress, making the experiment highly impractical. In contrast, seminal experiments have successfully demonstrated particle creation from the vacuum without moving a mirror, instead generating observable photons by modulating the electric boundary conditions of a superconducting cavity \cite{WILSONC.M2011Ootd} or using light in a superconducting metamaterial \cite{LahteenmakiPasi2013DCei}. However, these experiments rely on resonant enhancement to produce the time-dependent boundary conditions in the DCE, where a single quantum of a driving field is converted into two entangled photons via a parametric process. Such particle creation processes are equivalently described as a parametric coupling between static cavity modes. Consequently, they cannot be interpreted as observations of particle content arising from a non-perturbative change in the boundary conditions--one that significantly alters the photonic cavity's spatial mode profile as originally proposed by Moore \cite{MooreGeraldT.1970QTot} (in our case splitting the photonic cavity into two). In contrast, the authors of Ref.~\cite{BrownEricG.2015Wdim} proposed to use a rapidly slammed mirror in a photonic cavity to generate local particles associated with spatial sub-regions of the vacuum. Such an experiment would access vacuum particle creation from non-perturbative boundary condition changes. However, this has remained out of experimental reach due to the extremely high speed at which the mirror needs to be inserted in the cavity.

\begin{figure*}[t]
\begin{center}
\includegraphics[width=0.8\linewidth]{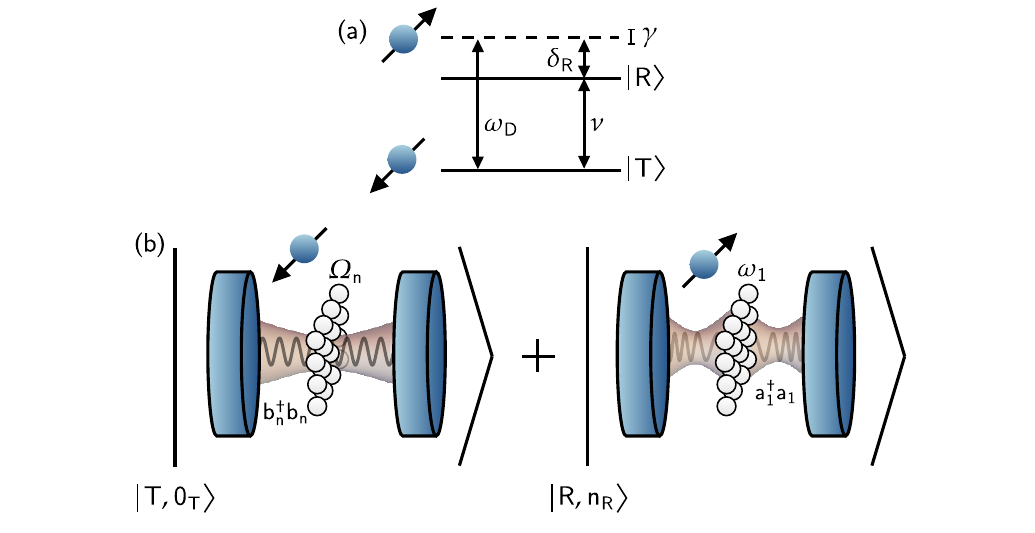}
    \caption{\label{n11combined} (a) When the reflectivity of the two-dimensional atomic array changes, it leads to particle creation from the vacuum. This vacuum-induced effect causes a frequency shift, denoted by $\delta_R$, in the transition frequency $\nu$ of the control atom. The shift occurs when the control atom is driven by an external field with drive frequency $\omega_D$. As we show in this work, the frequency shift can be much larger than the linewidth $\gamma$ of the control atom. (b) A quantum metasurface acts as a quantum-controlled mirror within a photonic cavity, where dynamical superpositions of reflective $(\ket{R})$ and transmissive $(\ket{T})$ states are created by driving the control Rydberg atom between its ground and excited state respectively. This dynamical superposition prepares superpositions of the cavity's boundary conditions, leading to observable frequency shifts of the control atom. These effects serve as a witness of particle content from entangled spatial sub-regions of the electromagnetic field vacuum, achievable without the need for classical rapid mirror motion. We note that while the array can be positioned at any place in the cavity, for practical realization we will consider it to be placed off-centre.}
\end{center}
\end{figure*}

In parallel, atomic arrays are at the forefront of modern quantum physics, with transformative applications in quantum computing, quantum simulation, quantum sensing and metrology \cite{JohanningMichael2009Qswc,EndresManuel2016Aaod,BrowaeysAntoine2020Mpwi,arya2024selectiveamplificationgravitationalwave,PRXQuantum.5.030316,PhysRevLett.133.013401,porras2008collective}. Specifically, sub-wavelength atomic arrays have been proposed as platforms for realising a quantum metasurface \cite{Bekenstein2020}, in which the atomic array can be toggled between dynamics of superpositions of transmissive $(\ket{T})$ and reflective $(\ket{R})$ states via control of the internal state of a single atom. Followed by proposals for employing them for atomic quantum operations \cite{PhysRevA.109.012613,antman2024atom}, their applicability in cavity set-ups is also receiving more attention, notably for their ability to provide confining mirrors in an optical cavity \cite{PhysRevA.111.053712}.

In this article, we propose a protocol to measure the particle content produced due to highly non-perturbative boundary condition changes using a quantum metasurface realised as a two-dimensional sub-wavelength atomic array \cite{Bekenstein2020}—that divides a photonic cavity. Motivated by earlier studies of quantum-superposed dynamics in cavity QED, where a quantum control is entangled with an optical cavity degree of freedom \cite{PhysRevA.56.3187,PhysRevA.72.022320,PhysRevLett.127.043603,Bekenstein2020,Sinha_2025} and studies of superposed vacuum creation effects in quantum field theory \cite{PhysRevD.102.085013,PhysRevD.102.045002}, we show that vacuum particle creation induced by strongly non-perturbative changes in boundary conditions produces a measurable frequency shift in the control atom of an atomic array, placing the effect within reach of realistic, though challenging experiments. As we show here, by mapping particle creation onto the control atom’s frequency shift, we sidestep the off-resonant suppression that occurs when the boundary condition changes on a timescale much shorter than the optical mode period (as in classical boundary-condition modulation). Mapping particle creation onto this frequency shift also removes the need for rapid modulation of a classical mirror’s reflectivity and instead enables quantum-controlled switching, while at the same time generating coherent dynamics of superpositions of distinct photonic QED vacua. Accessing this regime of cavity qed further enables the simulation of quantum effects arising due to the dynamics of superpositions of spacetimes, whereby the configuration of a source mass and its effect on fields \cite{PhysRevLett.129.181301, delaHametteAnne-Catherine2023Qrff} plays an analogous role to the quantum-controlled permittivity of the atomic array. 

In essence, observing the photon-induced frequency shift would constitute the first direct observation of vacuum particle creation in the sense of Moore’s original work \cite{MooreGeraldT.1970QTot}, i.e. driven by a non-perturbative change in the field’s boundary conditions that partitions the electromagnetic vacuum into spatial sub-regions.

\section{Sub-region particle content with a classical mirror}\label{subregpartcont}
We begin by outlining the theoretical framework for observing photon content from entangled spatial sub-regions of the vacuum with a rapidly introduced mirror, examined in Ref.~\cite{BrownEricG.2015Wdim}. Consider a photonic cavity of length  $L$, where the mode functions satisfying the Klein-Gordon equation with Dirichlet boundary conditions are given by \cite{BrownEricG.2015Wdim}:
\begin{equation}\label{modefunctions}
    U_n(x, t)=\frac{1}{\sqrt{L \Omega_n}} \sin \left(\frac{\pi n x}{L}\right) e^{-\mathrm{i} \Omega_n t},    
\end{equation}
where $\Omega_n = \frac{\pi n c}{L}$ are the resonance frequencies of the photonic cavity.
The corresponding eigenstates are Fock states of the cavity’s free Hamiltonian,  $\hat{H} = \sum_{n=1}^{\infty} \Omega_n \hat{b}_n^{\dagger} \hat{b}_n$, where  $\hat{b}_n^{\dagger}$  and  $\hat{b}_n$  are creation and annihilation operators for mode  $n $, satisfying commutation relations $[\hat{b}_n, \hat{b}_m^{\dagger} ] = \delta_{nm}$. Now, if a mirror is suddenly introduced at time  $t = 0$  within the cavity, it alters the boundary conditions and introduces time-dependence into the Hamiltonian \cite{BrownEricG.2015Wdim}:
\begin{equation}\label{mirrorham1}
    \hat{H}(t) =
\begin{cases}
\sum_n \Omega_n \hat{b}_n^{\dagger} \hat{b}_n & t < 0 \\
\sum_m \omega_m \hat{a}_m^{\dagger} \hat{a}_m + \bar{\omega}_m \hat{\bar{a}}_m^{\dagger} \hat{\bar{a}}_m & t \geq 0,
\end{cases}   
\end{equation}
where $\omega_m = \frac{\pi m}{r}$, $\bar{\omega}_m = \frac{\pi m}{\bar{r}}$  are the frequencies of the left and right sub-cavities of length $r$ and $\bar{r} = L - r$ respectively. The mode functions of the sub-cavities have the form Eq.\ \eqref{modefunctions}, with $L$ replaced by $r$, $\bar{r}$, respectively. Together, the left and right sub-cavity modes span the same space as the full cavity, which means that the full cavity modes can be expressed in terms of sub-cavity modes. This relationship is given by the Bogoliubov relation between  $\hat{b}_n$  and the sub-cavity modes $\hat{a}_j, \hat{\bar{a}}_j$ \cite{BrownEricG.2015Wdim}. In particular, the mode transformation for $\hat{a}_j, \hat{\bar{a}}_j $  takes the form,
\begin{equation}\label{bogoaaa}
\begin{split}
    \hat{a}_j &= \sum_{n=1}^{\infty} \left( \alpha_{jn} \hat{b}_n - \beta_{jn} \hat{b}_n^{\dagger} \right)\\
    \hat{\bar{a}}_j &= \sum_{n=1}^{\infty} \left( \bar{\alpha}_{jn} \hat{b}_n - \bar{\beta}_{jn} \hat{b}_n^{\dagger} \right),
\end{split}
\end{equation}
where the Bogoliubov coefficients are  $\alpha_{j n}=\frac{j(-1)^j}{\sqrt{n j} a \pi\left(n-\frac{j}{a}\right)} \sin (n \pi a)$  and  $\beta_{j n}=\frac{j(-1)^j}{\sqrt{n j} a \pi\left(n+\frac{j}{a}\right)} \sin (n \pi a)$ for the left sub-cavity, and $\bar{\alpha}_{j n}=\frac{-j}{\sqrt{n j} \bar{a} \pi\left(n-\frac{j}{\bar{a}}\right)} \sin (n \pi a)$  and  $\bar{\beta}_{j n}=\frac{-j}{\sqrt{n j} \bar{a} \pi\left(n+\frac{j}{\bar{a}}\right)} \sin (n \pi a)$ for the right sub-cavity ($a = \frac{r}{L}$, $\bar{a} = \frac{\bar{r}}{L}$ is the ratio between the cavity lengths of the left and right sub-cavities to the global cavity respectively). These Bogoliubov coefficients quantify the mixing between the original modes and the new modes due to the mirror’s sudden introduction. The nonzero  $\beta_{jn}, \bar{\beta}_{jn}$  coefficients signify local photon content in the global vaccum, as the vacuum state of the global cavity is no longer empty in terms of the sub-cavity modes.

The average number of sub-cavity photons in mode $i$, evaluated in the vacuum state of the original global cavity, is given by $\left\langle \hat{n}_i(t) \right\rangle = \sum_{n \geq 1} \left| \beta_{in} \right|^2$. For the extremal case where the mirror is placed at the midpoint of the cavity, such that the ratio of the sub-cavity length to the global cavity length is  $a = 0.5$ , the number of created photons can reach values as high as  $\langle \hat{n}_i(t) \rangle \simeq 0.05$. However, achieving this effect requires the spatial boundary conditions of the cavity to change non-perturbatively (in contrast to the experiments of Refs.~\cite{WILSONC.M2011Ootd,LahteenmakiPasi2013DCei}) on a timescale faster than the free evolution of the cavity modes. It was suggested in \cite{BrownEricG.2015Wdim} that a classical mirror with rapidly varying reflectivity could induce such fast and non-perturbative spatial boundary condition changes. Yet, this approach demands changes on the order of  $10^{-14} \; \mathrm{s}$, which is beyond current experimental capabilities due to the limitations of material response times \cite{responsetimes}. 

\section{The quantum-controlled photonic cavity}

We previously saw that in order to observe particle content from the vacuum due to non-perturbative boundary condition changes with
a classical metasurface, the transition from a transmissive to reflective metasurface would have to occur rapidly. In contrast, in this section we show how a quantum metasurface can make such particle content observable without the need for such rapid response times.

Our proposal places a sub-wavelength array of $N$ neutral atoms close to the end of a single-mode Fabry-Perot cavity. As before, we denote the size of the global cavity by $L$ and the left and right sub-cavities by $r$ and $\bar{r}$ respectively. We now assume that the atoms are situated close to the left mirror, such that the left sub-cavity is much smaller than the size of the global cavity, that is, $r \ll L$. Although the atoms could also be placed in the middle of the cavity (for which $r = \bar{r}$), we choose to work in this small sub-cavity regime in order to obtain a highly accurate analytical result, as well as maximise the frequency shift (in the $r \ll L$, the mode mis-match between the left sub-cavity mode and the global cavity mode functions is maximised leading to the most significant effect). As we demonstrate later, these results act as an order-of-magnitude estimate for other configurations, such as an array positioned close to the centre. We further note that for infinitely large $L$ the QED vacuum can function as the global cavity. The relevant internal states of the array atoms are denoted the ground state $\ket{g}$, excited state $\ket{e}$ and Rydberg state $\ket{r'}$. When placed in a sub-wavelength configuration, such a two-dimensional array of atoms naturally acts as a near-perfect mirror for photons tuned to its collective $g\leftrightarrow e$ dipole transition.  When a strong classical control field drives the $e \leftrightarrow r'$ transition, the array enters an EIT (electromagnetically-induced-transparency) regime: the destructive interference between the two excitation pathways opens a narrow transparency window, so the metasurface becomes almost loss-free and transmissive. The atom array therefore provides an implementation of a classical mirror in the cavity, which we considered in the previous section, depending on whether the EIT control beam is switched on or off. Implementing this sub-wavelength atom array in a cavity set-up remains the first step towards the implementation of our proposal.

To control the transmissive or reflective state of the atom array, we couple it to a neutral control atom. This was first proposed in Ref.~\cite{Bekenstein2020}.  By changing the state of the control atom, its interaction with the collective modes of the atom array tunes the array between transmissive and reflective states. When the control atom is excited to its own Rydberg state, the dipole-dipole (blockade) interaction shifts the $|r'\rangle$ level of every atom in the array, detuning the two-photon resonance and quenching EIT. As a result, any light injected into the cavity then sees the bare, highly reflective array, so the metasurface becomes a mirror for light resonant with the $g\leftrightarrow e$ dipole transition.  Preparing the control atom in a quantum superposition of ground and Rydberg states therefore entangles the array with the control atom and realises a coherent superposition of the two macroscopic optical responses---transparent (EIT-enabled) and reflective. In other words, the state of the control atom determines whether the atom array divides the global cavity into two sub-cavities or not.

We model the control atom as a qubit with control state $\ket{R}_c$ (for a reflective array) and $\ket{T}_c$ (for a transmissive array), as done in Ref.~\cite{Bekenstein2020}. Such a sub-wavelength atom array tunable between quantised transmissive and reflective states has been observed experimentally in a free-space set-up rather than a cavity set-up in Ref.~\cite{SrakaewKritsana2023Asaa}, albeit not yet extended to the coherent superpositions of Ref.~\cite{Bekenstein2020}. 

The total Hamiltonian of the array-control-atom system can be expressed as
\begin{equation}\label{Harray}
    \hat{H} = \hat{H}_\mathrm{array} + \hat{H}_\mathrm{free}+ \hat{U}_\mathrm{ryd},
\end{equation}
where $\hat{H}_\mathrm{array}$ is the many body Hamiltonian of the array (see Appendix~\ref{arrayresponse}), $\hat{H}_\mathrm{free}$ is the free Hamiltonian of the control atom, and $\hat{U}_\mathrm{ryd} = \sum_i \hat{V}_i = \sum_i V_{i c}|R\rangle_i\left\langle\left. R\right|_i \otimes \mid R\right\rangle_c\left\langle\left. R\right|_c\right.$ is the Rydberg-Rydberg interaction between the control atom and each atom in the array. This produces different dynamics of the array atoms conditioned on the quantum state of the control, essentially dynamics with or without the Rydberg interaction term. As a result, the permittivity of the atomic array, and therefore its reflectivity, produces a reflection coefficient dependent on the quantum state of the control. Previously, the permittivity of the array (which controls the reflectivity) was only computed when the control atom was in either the $\ket{R}_c$ or $\ket{T}_c$ state, inferring its extension to the case of a quantum-controlled permittivity \cite{Bekenstein2020}. In this work, we derive the reflectivity for exact dynamics of superposition states of $\ket{R}_c$ and $\ket{T}_c$, which is crucial for our proposal, and derive the relevant quantities, extending previous derivations. Henceforth, for ease of readability, we denote the extended collective mode of the many body state of the atomic array with the same quantum state as the control's quantum state. In this sense, one should be aware that the states $\ket{R}$ and $\ket{T}$ mathematically represent the full state of the control atom and the array on the joint Hilbert space $\mathcal{H}_\mathrm{control} \otimes \mathcal{H}_\mathrm{array}$ (and we have dropped the subscript ``$c$'' to indicate this). 

So far, we have not yet considered the optical field in the cavity. When the atom array is tuned to the transmissive state $\ket{T}$, the global cavity supports standing modes of the optical field, which we denote by
\begin{align}\label{HT111}
    \hat H_{T} = \sum_n \Omega_n \hat b_n^\dag \hat b_n,
\end{align}
where $\Omega_n$ is the frequency of the supported global modes, $\hat{b}_n$, in the cavity. We show in  Appendix~\ref{reflappdev} that when the atom array is reflective to a set of frequencies in some reflectivity bandwidth $\Delta$, the Hamiltonian describes now two sub-cavities, where the left sub-cavity confines  the mode we denote $\hat{a}_1$ with frequency $\omega_1$, and the right sub-cavity, which confines a set of modes $\hat{\bar{a}}_k$ with frequencies $\bar{\omega}_k$; explicitly this Hamiltonian reads 
\begin{align}\label{HRRR}
\hat{H}_R  = \sum_n \Omega_n \hat{b}_n^{\dagger} \hat{b}_n + \omega_1\hat{a}_1^{\dagger}\hat{a}_1 + \sum_k \bar{\omega}_k\hat{\bar{a}}_k^{\dagger}\hat{\bar{a}}_k, 
\end{align}
where as in the previous section  $\hat{b}_n$ refer to the global cavity modes of frequency $\Omega_n$. This expression is valid in the regime for which the global cavity length is much larger than the smaller sub-cavity length, $r \ll L$. In addition, we require the array to be reflective in a narrow frequency range, specifically, we require $\Delta$, to be much smaller than the sub-cavity mode frequency spacing, \textit{i.e} free spectral range (FSR): $\Delta \ll \omega_1$.

We now consider the case where the control atom is in a superposition of reflective and transmissive states, that is, $\ket{\psi} = (c_R\ket{R} + c_T\ket{T})$. By linearity, the cavity Hamiltonian conditioned on the control’s quantum state is  
\begin{equation}\label{hallterms}
        \hat H_\mathrm{cav} = \hat H_T \otimes | T \rangle\langle T | + \h H_R \otimes | R \rangle\langle R|, 
\end{equation}
which we derive through a quantum-control state-dependent interaction of the atomic array with the quantised field modes, starting from the microscopic QED interaction $\hat{H}_{\mathrm{int}}=\hat{\mathbf{E}}(\mathbf{x}) \cdot \hat{\mathbf{P}}(\mathbf{x})$ in Appendix~\ref{QCHS}. Here, $\hat{H}_T$ is the global cavity Hamiltonian of Eq.~\eqref{HT111} and it appears in $\hat{H}_\mathrm{cav}$ conditioned on the control being in the transmissive state.  $\hat{H}_R$ represents the reflective Hamiltonian, given in Eq.~\eqref{HRRR}, and it appears in $\hat{H}_\mathrm{cav}$ conditioned on the control being in the reflective state. We re-iterate that this is valid in the regime for which the global cavity length is much larger than the sub-cavity length $r  \ll L$, and the array is reflective in a narrow frequency range, $\Delta \ll \omega_1$.

Finally, we consider the mechanism by which the control atom is driven from the transmissive $\ket{T}$ state to the reflective $\ket{R}$ state. In the full dynamics, the control atom cannot be switched from $\ket{T}$ to $\ket{R}$ instantaneously; the corresponding drive Hamiltonian between such orthogonal states reads: 
\begin{align}
\hat{H}_{\text {switch }}=g\left(|T\rangle\langle R| e^{i\omega_D t}+\text { h.c }\right).
\end{align}
Here, $g$ corresponds to the Rabi coupling strength of the control atom with an incident drive of frequency $\omega_D$. The term $\h H_\mathrm{switch}$ thus describes how the control atom can drive transitions between the transmissive and reflective cavity states. The full Hamiltonian of the atom, cavity and control atom is $\hat{H}  = \hat{H}_\mathrm{cav} + \hat{H}_\mathrm{switch} + \hat{H}_\mathrm{free}$, where the free Hamiltonian of the control atom can be explicitly written as
\begin{equation}
    \begin{gathered}
\hat{H}_{\text {free }}=\frac{\hbar \nu}{2}(|R\rangle\langle R|-|T\rangle\langle T|),
\end{gathered}
\end{equation}
where the transition frequency between the two relevant states is denoted $\nu$.

\section{Observing sub-region particle content with a quantum metasurface}\label{sub222}
The quantum-controlled coupling of the electromagnetic field energy to the control atom causes the vacuum state of the field to contain non-zero particle content. This, in turn, re-normalises the energy levels of the control atom. Our goal is to derive an analytic estimate of this frequency shift of the control atom due to vacuum particle content, arising from the change of the array’s reflectivity, i.e.\ the appearance of a mirror in the cavity. 

The effect of the local particle content of the cavity vacuum can be best understood by considering the form of the free Hamiltonian of the atom and the cavity, $\hat{H}_{\text {cav }} + \hat{H}_{\text {free }}$. This expression can be reorganised to emphasize how the control system’s energy levels depend on the metasurface response and thus on the modes of the photonic cavity,
\begin{equation}\label{fullshifth2}
\begin{split}
     &\h H_\mrm{cav} + \h H_\mrm{free} 
    \nonumber \\
    &= \bigg( \h H_R + \frac{\hslash\nu}{2} \bigg) \otimes | R \rangle\langle R | + \bigg( \h H_T - \frac{\hslash\nu}{2} \bigg) \otimes | T \rangle\langle T |, \nonumber
\end{split}
\end{equation}
where $\hat{H}_R=\hat{H}_T +  \omega_1 \hat{a}_1^{\dagger} \hat{a}_1  +  \sum_k \bar{\omega}_k \hat{a}_k^{\dagger} \hat{a}_k$, as in Eq.~\eqref{HRRR}, where it becomes clear that the energy levels of the control atom are renormalised by the distinct energy contributions from the global and sub-cavities, described by $\hat{H}_T$ and $\hat{H}_R$ respectively. The frequency shift depicted in Fig.~\ref{n11combined} ($\delta_R$), can thus be interpreted as a shift in the atom’s resonance frequency, which is caused by the differing energy states of these two cavity configurations. As we will discuss further, this shift arises due to the sub-cavity having non-zero particle content for the global cavity vacuum state in the fast-switching regime (boundary condition changes on time-scales comparable to or much faster than the frequency of the optical mode $\tau_g \gtrapprox \frac{1}{\omega_1}$), and non-zero particle creation in the slow switching regime (timescales much smaller than the free-evolution of the optical modes). 

In order to further understand the frequency shift, we convert $\mathrm{H}_\mathrm{switch}$ into the interaction picture with respect to $\hat{H}_{\mathrm{cav}}~+~\hat{H}_{\mathrm{free}}$, where the unitary evolution of the system, up to first order in  $g$ is
\begin{equation}
    \hat U_I^{(1)} = \hat{\mathds{I}} - i g \int_0^t \mathrm{d}t^{\prime} \left( e^{i\hat{H}_R t^{\prime}} |R\rangle\langle T| e^{-i ( \hat{H}_T + \delta ) t^{\prime} } + \mathrm{h.c} \right),
\end{equation}
where $\delta = \nu - \omega_D$ is the detuning between the laser frequency and the control atom transition. Starting from an initial vacuum state of the global cavity $\ket{0_T}$ and the state of the control atom for which the metasurface is transmissive  $|T\rangle$, the probability to measure the control atom in the reflective state  $|R\rangle$  becomes (considering a weak drive $g \ll \omega_1$, where we recall that $\omega_1$ is the frequency of the fundamental mode of the sub-cavity):
\begin{equation}\label{PR1234}
\begin{split}
    P_R &= 4 (gt)^2 \left\langle \mathrm{sinc}^2 \bigg( \frac{1}{2} \Big( \delta + \omega_1 \h a_1^\dagger \h a_1 + \sum_n \Omega_n \hat{b}_n^\dagger  \hat{b}_n \Big) t \bigg) \right\rangle \\ 
    &\simeq 4 (gt)^2 \mathrm{sinc}^2 \bigg( \frac{1}{2} \Big( \delta + \delta_R \Big) t \bigg),
\end{split}
\end{equation}
where the expectation value is taken with respect to the field state after the switch, and $\hat{a}_1$, $\hat{b}_n$ represent the fundamental sub-cavity and global mode operators respectively, and $\delta_R$ is an analytical estimate of the frequency shift due to the particle content created from a time-dependent boundary condition change. In order to observe particle content, we now focus on two distinct regimes where the atom array is switched from transmissive to reflective.

\subsection{Slow-switching}\label{sss1}

In the most experimentally relevant regime where the array is switched from transmissive to reflective on a timescale comparable to the excited-state linewidth ($\Gamma_e$) of the array atoms, with $\tau_e \sim 1/\Gamma_e$, the photon content arises from the time-dependent reflectivity of the array. As the Hamiltonian changes from $\hat{H}_T \rightarrow \hat{H}_R$, the reflectivity coefficient varies in time and the initial vacuum state of the field is no longer the vacuum of the instantaneous Hamiltonian. This deformation of the vacuum, driven by a highly non-perturbative change in the boundary conditions, produces spatially localised photons from the vacuum, with a photon number of order $n \sim (\Gamma_e / \omega_1)^2$ (see Appendix~\ref{multinon}). In this regime, the sub-cavity field evolves into $\ket{\beta_R} = \ket{0} + \frac{\Gamma_e}{\omega_1}\ket{1} + \; \mathcal{O}((\frac{\Gamma_e}{\omega_1})^2) $, which is a coherent state of amplitude $\frac{\Gamma_e}{\omega_1}$. This change of state arises due to the time-dependent boundary as the array switches from the EIT Dark state to a highly reflective state, as outlined in Appendix.~\ref{timesec}, leading to the non-zero particle content, 
\begin{equation}
    \bra{\beta_R}\hat{a}_1^{\dagger} \hat{a}_1\ket{\beta_R} = |\beta_R|^2 \geq 0,
\end{equation}
Importantly, the relevant particle content is controlled by the finite switching rate (or ramp time) and is vanishing in the adiabatic (infinitely slow switching) limit—relative to the sudden-switch value, as derived in Appendix~\ref{multinon}. 

In summary, this protocol constitutes a novel approach to observe vacuum-particle production due to non-perturbative boundary condition changes. This is achievable with this system due to the scaling enhancement by the factor $\omega_1$ for measuring particle content compared to the noise threshold. This can be extrapolated from Eq.~\eqref{PR1234} that clearly provides an $\omega_1$ enhancement factor as compared to the classical mirror modulation of Ref.~\cite{PhysRevA.99.053833} for witnessing vacuum particle production (we also go further beyond Ref.~\cite{PhysRevA.99.053833} and other proposals for witnessing parametric particle production by working in the regime of highly non-perturbative boundary condition changes as we dicuss further in Section~\ref{Disc1}). To see this enhancement explicitly, for the control atom's linewidth $\gamma$ - the vacuum frequency shift compared to the linewidth is $\frac{\omega_1\left|\beta_R\right|^2}{\gamma}$.

\subsection{Fast-switching}\label{fastsss1}
In the fast switching regime (which requires improvements beyond the current state-of-the-art), the expectation value in Eq.~\eqref{PR1234} is taken with respect to $\ket{0_T}$, as the field state remains unchanged before and after the switch, and $\delta_R = \sum_n \omega_1\left|\beta_{1, n}\right|^2$ is an analytical estimate of the frequency shift due to the sub-region particle content in this fast switching regime (see Appendix~\ref{analytical111} for a derivation in the rotating frame of $\hat{H}_R$, after expanding sub-cavity modes in terms of global cavity modes with a Bogoliubov transformation or multi-mode squeezing operation \cite{MaXin1990Msoa}). We present the derivation of the above perturbative expression for the transition probability in Appendix~\ref{perturblim}. From Eq.\ \eqref{PR1234}, it is apparent that the presence of photons in the sub-cavity with respect to the global cavity vacuum state, 
\begin{equation}\label{analyticest}
    \bra{0_T}\hat{a}_1^{\dagger} \hat{a}_1\ket{0_T} = \sum_n|\beta_{1n}|^2 \geq 0,    
\end{equation}
produces a frequency shift of the control atom. This analytical expression for the frequency shift is exact in the $r \ll L$ regime, while providing an order-of-magnitude estimate when the atomic array is positioned elsewhere in the cavity. This particle content in Eq.~\eqref{analyticest} that produces the characteristic frequency shift corresponds to the case of ideal rapid switching, which requires Purcell-enhanced-linewidths several orders of magnitude beyond state-of-the-art \cite{WangJian2025UHSR} for the exact protocol described in this section to work (as we detail further in section~\ref{expconsider1}). This formula describes a frequency-shift that peaks for $a \rightarrow 0$, with a $\mathrm{sinc}$ function decay to zero as $a \rightarrow 1$, as suggested by the expressions for the Bogoliubov co-efficients outlined in Section.~\ref{sub222}. The asymmetry in the frequency shift arises because the mirror only reflects a single frequency - higher order modes have a comparatively negligible contribution, leaving only the contribution from the fundamental mode which has this characteristic curve.

\begin{table*}[t]
\begin{ruledtabular}
\begin{tabular}{lcccc}
Switching regime & Required parameters &Switching parameter &  Switching time-scale & Measured Frequency Shift \\
\hline
\text Slow non-parametric & $\frac{\Omega_p^2}{\Gamma_eV} \ll 1$  & $\Gamma_e$ &  $ 10 \; \mathrm{ns}$ \; ($1 \; \mathrm{ns}$)  & $0.1 \; \mathrm{Hz}$  ($10^{-16}$)
\\
\text Slow non-parametric enhanced & $\frac{\Omega_p^2}{\Gamma_eV} \ll 1$  & $\Gamma_e$ &  $ 0.001 \; \mathrm{ns}$ \; ($1 \; \mathrm{ns}$)  & $100 \; \mathrm{Hz}$ ($10^{-13}$)
\\
\text Slow parametric & $\frac{\Omega_p^2}{\Gamma_eV} \gg 1$ & $\Omega_p$  &  $\mathrm{ps}$ \; ($\mathrm{ns}$) & $1 \; \mathrm{Hz}$ ($10^{-15}$)
\\
\text Fast non-parametric & $\frac{\Omega_p^2}{\Gamma_eV} \ll 1$ & $\Gamma_e$  & $\mathrm{fs}$ \; ($0.1 \; \mathrm{ns}$)  & $10^{11} \;\mathrm{Hz}$ ($10^{-4}$)
\\
\end{tabular}
\end{ruledtabular}
\caption{\label{tabular1111} Table of parameters for switching time-scales of the atomic array's reflectivity needed to produce the particle-content-induced frequency shift (of the stated order of magnitude) in the final column, with comparison to state-of-the-art switching times in brackets in the fourth column. Standard atom linewidths are on the order of $\Gamma_e \sim 10 \; \mathrm{MHz}$, with an assumed two order of magnitude enhancement of $\Gamma_e \sim 1000 \; \mathrm{MHz}$. Extreme Purcell-enhanced linewdiths are on the order of $\Gamma_e \sim \mathrm{THz}$ as stated in the main text. The EIT control beam coupling strength for standard array parameters (that represents rows 1,2 and 4) is on the order of $\Omega_p \sim \mathrm{MHz}$. For strong EIT control beam coupling strengths $\Omega_p \sim 10 \; \mathrm{GHz}$, one can undertake slow switching parametric particle creation as highlighted in the third row, while typical $\mathrm{MHz}$ strengths would suppress it to Hz-scale frequency shifts. The different regimes can be accessed by initialising the control atom in its ground state, and driving transitions to the excited state - the relative strength of the Rydberg interaction energy ($V$) in comparison to the EIT linewidth (second column) allows access to the distinct regimes. The last column quantifies the frequency shift and converts it into a fractional frequency shift.}
\end{table*}

\subsection{Extraction Protocol}
The following protocol can be used to verify the predicted vacuum-induced frequency shift of the control atom in either the fast or slow switching regimes with the shifted frequency as outlined in Eq.~\eqref{PR1234}. Firstly, the control atom is prepared in the transmissive state $\ket{T}$. Next, the control atom is pumped on-resonance with the control atom's un-renormalised transition frequency $\nu$. In this case, due to the frequency shift from the presence of vacuum-photons, there will be a significantly reduced probability for the control atom to flip to the reflective state $\ket{R}$. If the smaller sub-cavity of length $r \ll L$ is then pumped on-resonance with sub-cavity mode $\hat{a}_1$, the typical Lorentzian peak in the power spectrum of the output mode (at the frequency $\omega_1$ of the mode $\hat{a}_1$) will not be observed. However, if the sub-cavity is pumped on-resonance with the re-normalised transition frequency $\nu' = \nu + \sum_n \omega_1\left|\beta_{1, n}\right|^2$ (or $\nu' = \nu + \omega_1\left|\beta_{R}\right|^2$ in the slow switching case), the array will flip to the reflective state (due to incident photons to the cavity being resonant with the $\ket{g} \rightarrow \ket{e}$ transition for the control atom), and the output power spectrum will have a peak at $\omega_1$, as expected given that the reflective Hamiltonian $\hat{H}_R$ forms standing waves at the frequency $\omega_1$. This average intensity will be suppressed by $g^2/(g^2+ \delta^2)$ for a drive detuned from the transition frequency by $\delta$, as demonstrated in Appendix~\ref{nonperturb1} (the average reflected number of photons will be weighted by the probability for the array to be in the reflective state). For a drive detuned by $\delta \sim 0.1 \; \mathrm{Hz}$, and Rabi coupling strengths as weak as $g \sim 10^{-14} \; \omega_1$, the suppression of the average intensity is by a factor of $1 - 10^{-4}$ compared to the intensity in the sub-cavity on-resonance. The dominant contribution to all particle-content-induced frequency shifts comes from the left (and smaller) sub-cavity. The off-resonant nature of the drive Hamiltonian with the cavity frequency ensures the photon content produced can only be explained by the boundary condition change rather than photons pumped into the cavity. 

If set-up enables direct access to the control atom, the frequency shift can further be measured through measurements on the control atom directly. We further note that frequency shifts smaller than the linewidth can be resolved in this manner through many repititions of the experiment, but this will require a large number of runs of the experiment (a single-shot Rabi drive and measurement of the cavity power spectrum will generally not be sufficient) - assuming $\sqrt{N}$ precision, then $N = 10^{10}$ iterations will give down to sub-$\mathrm{Hz}$ precision. We emphasize that this regime involves standard EIT probe beam coupling strengths and linewidth limited switching times. By removing the 1 second fluorescence imaging step in Ref.~\cite{SrakaewKritsana2023Asaa} and making measurements just on the control atom the time-scale to observe the effect can be reduced - if we use  $\mathrm{ms}$ scale time-scales for the control atom preparation and measurement (with the number of measurements before a re-preparation is required then being limited by the motional ground state lifetime of the array atoms). With Purcell enhancement, or strontium atoms (which permit even larger linewidths \cite{MaugerS2007SosR}), or alternating cooling and measurement procedures this time-scale can also be independently reduced (or alternatively with the full second-scale procedure one order of magnitude Purcell-enhanced linewidths would require an experiment run-time of single years or months). These numbers indicate that this would be a significantly challenging experiment with existing atom array technology - although a plausible quantum-enhanced pathway to the non-perturbative dynamical Casimir effect.

\begin{figure*}[t]
    \centering
    \includegraphics[width=\linewidth]{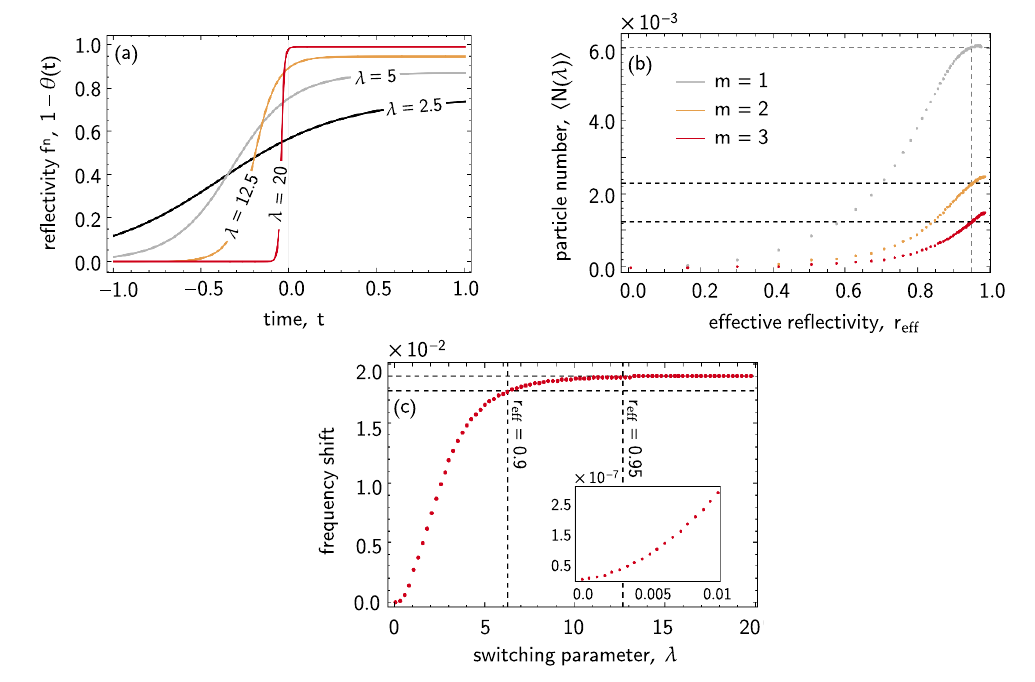}
    \caption{(a) Plot of the switching profile $r(t) = 1 - \theta(t)$ as a function of time, where $r = 1$ corresponds to perfect reflection. The orange curve depicts an effective reflectivity $r_\mrm{eff} = 0.95$. (b) Plot of the particle content of the $m$th sub-cavity mode in the global vacuum state as a function of the effective reflectivity $r_\mrm{eff}$. We have plotted this difference for a cavity with walls at $x = \pm a/2$ with $a = 1$. The dashed lines correspond to $r_\mrm{eff} = 0.95$. (c) Plot of the estimated frequency shift re-normalised to the sub-cavity frequency $\omega_1$, $\frac{\delta_R}{\omega_1} = \sum_n | \beta_{1,n} |^2$ obtained within our toy model, as a function of the switching parameter $\lambda$. The inset plots the frequency shift in the slow-switching regime corresponding to low effective reflectivities $r_\mrm{eff} \sim 10^{-4}$.  Recall that $\lambda$ serves a dual purpose: it determines both the proximity of the mirror’s final state to perfect reflectivity and the speed at which the mirror is activated.}
    \label{fig:suppression}
\end{figure*}

\section{Observable Frequency-Shift Magnitude and Scaling}\label{expconsider1}
\subsection{Array Parameters}
For the protocol described in Section.~\ref{sub222} to lead to the successful observation of such vacuum particle content, several experimental requirements must  be met. Here we discuss the parameter requirements of the atomic array system in order to witness the discussed effects in the fast and in the slow switching regimes.

In order to observe either the vacuum’s unperturbed particle content or slow-switching particle creation from strongly non-perturbative boundary changes, our protocol requires that the control atom acts as the quantum degree of freedom setting the cavity boundary condition. In order to achieve this, it is crucial that the Rydberg blockade of the control atom reaches each atom in the array, such that driving the control atom into its Rydberg state switches off the transparency through the Rydberg blockade as in Ref.~\cite{Bekenstein2020}. Although switching the control atom between states takes $O(1/g)$ time, in the reflective branch of the superposition the energy levels of the array atoms shift within the light-crossing time. After this, the atom array’s reflectivity switches on, changing from the EIT dark-state transmissivity to a reflective state determined by the array properties—notably the excited state linewidth $\Gamma_e$ and the EIT control-beam coupling strength $\Omega_p$. We classify four regimes of relevance with varying experimental feasibility as summarised in Table~\ref{tabular1111}. 

To capture the different physical behaviours of the system, we analyse three distinct parameter regimes in a full time-domain treatment of the reflectivity switch, presented in Appendix~\ref{arrayresponse}. We firstly focus here on the first regime, which is based on the standard Rydberg blockade mechanism for switching on the reflectivity. This is the same physical process used in the experiment of Ref.~\cite{SrakaewKritsana2023Asaa} where strong interactions between Rydberg excitations prevent multiple atoms enabled the switch of the collective mode of the array from transmissive to reflective correlated with the state of the quantum-control.

In the limit where the Rydberg interaction strength is much larger than the effective EIT linewidth, $\Omega_p^2/\Gamma_r$ (with $\Omega_p$ the control-beam coupling strength and $\Gamma_r$ the Rydberg decay rate), we find that the reflectivity rises on a timescale set by the excited-state linewidth. As a result, the system operates in the slow-switching regime discussed in Section~\ref{sss1}, and is accessible with standard array parameters involving MHz EIT probe beam coupling strengths and linewidth limited switching rates.

This slow switching leads to a small but finite frequency shift of the optical mode, which scales as $(\Gamma_e/\omega_1)^2$ (see Appendix~\ref{multinon}), where $\Gamma_e$ is the excited-state linewidth and $\omega_1$ is the mode frequency. The corresponding particle production in this regime is shown in the first row of Table~\ref{tabular1111}. Importantly, this shift can be significantly enhanced using Purcell-enhancement, bringing it into the kilohertz range (as indicated in the second line). The resulting timescale defines the central operating regime of this work and is crucial for enabling non-perturbative vacuum particle creation to be observed with near-term experimental technology.

The second regime of slow-switching parametric particle production, as given in Table~\ref{tabular1111}, line three, is that of small Rydberg interaction strengths compared to the EIT linewidth. We find that fast oscillations of low effective reflectivities (see Appendix~\ref{arrayresponse} Figure~\ref{ref}), will enable the switching time to oscillate on the time-scale of the EIT control beam coupling strength $\Omega_p$, which for typical EIT linewidths corresponds to a similiar MHz switching rate. While accessible with near-term set-ups, this regime enables parametric particle production in the global cavity akin to the experiments of Refs.~\cite{WILSONC.M2011Ootd,LahteenmakiPasi2013DCei}, i.e. the boundary condition change is perturbative reducing to parametric squeezing operation affecting the global cavity modes, as outlined in Appendix~\ref{beam1}. 

The final regime given in Table~\ref{tabular1111} corresponds to the fast switching regime described in Section~\ref{fastsss1}, i.e. switching time-scales sufficiently fast compared to the time-scale for the free-evolution of the fundamental sub-cavity mode ($2.5 \; \mathrm{fs}$ for $\frac{\omega_1}{2\pi} \sim 400 \; \mathrm{THz}$). The first of these non-adiabatic regimes would correspond to a Purcell-enhanced linewidth \cite{WangJian2025UHSR} by several orders of magnitude for the standard linewidth-limited switching of the first regime discussed above. In this way, if the Rydberg-blockade-induced energy shift is much larger than the EIT linewidth, if the array atom's $\ket{g} \rightarrow \ket{e}$ transition has a radiatively enhanced linewidth by several orders of magnitude (as in the Purcell effect), the switching time-scale can approach the femto-second regime, as Purcell broadening of the linewidth will enhance the switching rate (see Appendix~\ref{timesec}). In this case, the protocol from Section~\ref{sub222} can be used to extract the frequency shift, but will require an additional photonic cavity coupled to the array atoms (achievable with orthogonal modes) to radiatively enhance their linewidths. However, achieving the enhanced time-scales this fast is out of reach of near-term technology. We emphasize that these final two regimes are significantly less practical than the first regime, as they require yet unobtained EIT probe coupling strengths, or extreme purcell enhanced linewidths.

In order to quantitatively assess the magnitude of the reported particle-content-induced frequency shift in both the fast and slow-switching regimes, we now compare the magnitude of the frequency shift to the linewidth of the control atom. We assume a fundamental sub-cavity mode frequency on the order of $400 \; \mathrm{THz}$ normalised to a control atom linewidth on the order of $10 \; \mathrm{kHz}$ \cite{Branden_2010}. Our results demonstrate that for ideal, fast-switching (given in the last row of Table~\ref{tabular1111}), the shift in the transition frequency is on the order of $2.8 \%$ of the total transition frequency, but over 11 orders of magnitude larger than the linewidth (due to the Bogoliubov transformation contributing a frequency shift that is approximately $7 \%$ of the photonic cavity frequency, see Appendix~\ref{analytical111} for more details). The numbers for the frequency shift above are computed for the case of ideal rapid switching (non-adiabatic boundary condition changes), which can be accessed in one of two central ways as summarised in Table.~\ref{tabular1111}. As displayed, this large frequency shift requires either a significantly Purcell-enhanced linewidth for the atoms in the atomic array (specifically by four orders of magnitude, which is three orders of magnitude beyond state of the art for Purcell-enhanced linewidths \cite{WangJian2025UHSR}). In this case, the array's response time-scale is set by the EIT control beam strength $\Omega_p$, leading to slow switching particle production. Alternatively, implementation can be completed in the fast-switching regime with the development of the quantum metasurface at microwave frequencies with superconducting artificial atoms, where such systems already have the parameters required for fast switching \cite{FriskKockumAnton2019Ucbl}, or the development of a quantum metasurface at mm-wavelengths, which would require appropriate transduction for the read-out as proposed in Ref.~\cite{KumarAishwarya2023Qmwt}.

In the slow switching regime, corresponding to the first two parameter regimes discussed in this section (and given in the first three rows of Table~\ref{tabular1111}), we show that the particle content is suppressed by a factor proportional to the ratio $(\Gamma_e/\omega_1)^2$ (where $\Gamma_e$ is time-scale for the reflectivity rise simulated in Appendix~\ref{timesec}), which corresponds to a frequency shift approximately an order of magnitude larger than the linewidth. This corroborates a simple toy model for a slowly switched mirror that yields a power law suppression of the produced particle content (see Fig.\ \ref{fig:suppression} and discussion below). More generally, the particle content created being proportional to $(\Gamma_e/\omega_1)^2$, once mapped onto a frequency shift on the control as per Eq.~\eqref{fullshifth2}, gives an $\omega_1$ factor enhancement of the signal above the linewidth of the control atom. This corresponds to an $\omega_1$ factor scaling enhancement in the feasibility of witnessing such vacuum particle creation effects compared to the classical mirror modulation of Ref.~\cite{PhysRevA.99.053833}. This analysis demonstrates how the quantum-controlled mirror can provide a quantum advantage over classical boundary condition changes for vacuum photon creation.  

\subsection{Atomic Species}
We now consider the atom species required to fulfill the requirements just discussed. The numbers used to compute the change in the transition frequency are adapted from a state-of-art experimental system of $^{87}\mathrm{Rb}$  sub-wavelength atom array that has potential for implementation of our proposed scheme \cite{SrakaewKritsana2023Asaa}, or alternatively with $\mathrm{Yb}$ atoms \cite{ShahmoonEphraim2017CRiL}. Specifically, we consider the following energy levels: $\ket{g} =\left|5 \mathrm{~S}_{1 / 2}, F=2, m_F=-2\right\rangle$ as the ground state and $\ket{e} = |5 \mathrm{P}_{3 / 2}, F=3, m_F=-3\rangle$ as the excited state, resonating at $\omega_1/2\pi = 400 \; \mathrm{THz}$. Accessing a fundamental mode of this frequency will require a cavity of order sub-micron length, specifically $375 \; \mathrm{nm}$ long. While such a small cavity will increase the resonance line width by a factor proportional to $\frac{1}{r}$, this will not substantially reduce the power spectrum for such small resonators. For lattice spacing $a \simeq 0.2\lambda_1$, where $\lambda_1$ is the wavelength of the fundamental sub-cavity mode, (also achievable with $\mathrm{Yb}$ atoms), corresponding to $150 \; \mathrm{nm}$, the array achieves perfect reflectivity \cite{ShahmoonEphraim2017CRiL}, making it reflective to modes near $\omega_1$. For sub-cavity linewidth $\kappa \ll \omega_1$, this frequency would be the only supported mode, allowing the atomic array to serve as one end of a sub-cavity, similar to that proposed in Ref.~\cite{PhysRevA.111.053712}. We consider the Rydberg control atom to also be a $^{87}\mathrm{Rb}$ atom, but consider the ground state of the ancilla control to be $\left|g^{\prime}\right\rangle = \left|5 \mathrm{~S}_{1 / 2}, F=1, m_F=-1\right\rangle$, with Rydberg state $\left|r_P\right\rangle = \left|44 \mathrm{P}_{3 / 2}, m_J=3 / 2\right\rangle$, which has a transition at $\nu/2\pi = 1000 \; \mathrm{THz}$, with achievable Rabi drive strengths of $g/2\pi \sim 1 \; \mathrm{MHz}$ \cite{SrakaewKritsana2023Asaa}, and assume an ancilla linewidth on the order of $\gamma/2\pi \simeq 10 \; \mathrm{kHz}$ \cite{Branden_2010}. Therefore, for this arrangement of a two-dimensional atomic array, if the array is $\frac{\lambda_1}{2} \simeq 375 \; \mathrm{nm}$ from one end of the cavity producing a sub-cavity with frequency $\omega_1$, the ideal frequency shift of the control atom will be $\delta_R \sim \mathrm{THz}$ (also corresponding to that resolvable with ultra-fast optics), with the fast oscillation of low reflectivities and slow-switching case an order of magnitude larger than the linewidth. The main limiting factor in extending the switchable mirror demonstrated in Ref.~\cite{SrakaewKritsana2023Asaa} to the formation of coherent dynamics of quantum superpositions of transmissive and reflective states is increasing the decay time of the control  atom 
from $T_1 \simeq 27 \; \mathrm{\mu s}$ demonstrated in Ref.~\cite{SrakaewKritsana2023Asaa} to become on the order of $T_1 \simeq 100 \; \mathrm{\mu s}$, as demonstrated in Ref.~\cite{PhysRevLett.123.170503}.

We further require that the control atom be able to be driven between the ground and excited state in a coherent manner. In order to achieve this, for the control-atom parameters considered here, and practical optical cavity $Q$-factors $\kappa$ on the order of $10^8$ or larger, the co-operativity is above unity for coherent oscillations. For example, using $\frac{g}{2\pi} \sim 10^6 \; \mathrm{Hz}$, $\frac{\kappa}{2\pi} \sim 10^8 $, and $\frac{\gamma}{2\pi} \sim 10^4 \; \mathrm{Hz}$ (where $\gamma$ is the linewidth of the control atom), gives well above unity for the Rabi-drive enhanced co-operativity, allowing coherent Rabi oscillations for the control atom. Moreover, such a small decay rate {$\kappa$} ensures that photons in the reflective branch of the superposition survive long enough to produce a measurable frequency shift.

\section{Noise and imperfections}

Two main sources of imperfections are thermal motion of the atomic array, and imperfect reflectivity. Regarding the former, assuming an atomic array that has a collective motional DoF cooled to the quantum ground state, the magnitude of thermal motion of the array will be on the order of its zero-point motion \cite{optoarray}. In sub-wavelength atom arrays the Lambe-Dicke parameter $\eta$ quantifies the spatial spread of an individual atom \cite{eltohfa2025effectsfinitetrappingdecay}. For example, the standard deviation of position divided by $\lambda_0$ (wave-length of the incident light) is equal to $\eta/2\pi$ for the vibrational state. It is given by $\eta=k_0 \sqrt{\hbar/(2m\omega_\mrm{t})}$, where $\omega_\mrm{t}$ is the trap frequency, which is typically on the order of several hundred kHz, and $k_0$ is the wavenumber of the incident light. To reflect optical light as we consider here, these parameters imply a positional spread on the order of $10 \; \mathrm{nm}$, much smaller than the cavity length. 

The reflectivity of the array is not perfect. To account for this, we utilise a toy model in Appendix~\ref{app:E} (see also Ref.\ \cite{BrownEricG.2015Sasc}) to compute the particle content produced by a smoothly switched reflective mirror that divides a Dirichlet cavity. Importantly, as the dominant particle content for the single frequency mirror in the main text is due to the fundamental sub-cavity mode, we can apply the corrections for the imperfect reflectivity for the fundamental sub-cavity mode to the single frequency mirror used in the main text. The switching of the mirror is determined by a one-parameter family of time-dependent profiles, $\theta (t) = (2/\pi)\tan^{-1} ( ( 1 + e^{-\lambda t} )/( \lambda))$. This choice of $\theta(t)$ is such that solutions to the cavity Klein-Gordon equation smoothly interpolate between transmissive in the asymptotic past, to partially (i.e.\ imperfectly) reflective in the asymptotic future. The ``effective reflectivity'' $r_\mathrm{eff} = 1 - \theta(\infty)$ (i.e.\ the closeness of the mirror's end state to a Dirichlet boundary) is controlled by the parameter $\lambda$, which also quantifies how quickly the mirror is switched on, see Fig.\ \ref{fig:suppression}(a). As shown in Fig.\ \ref{fig:suppression}, both the particle content of the sub-cavity modes and the estimated frequency shift induced on the control atom are only slightly suppressed (see Fig.\ \ref{fig:suppression}(b)-(c)) for experimentally expected reflectivities $r_\mrm{eff} \sim 0.95$ \cite{Bekenstein2020}. Even in the slow-switching regime for which $\lambda \sim 10^{-3}$ ($r_\mrm{eff} \sim 10^{-4}$), corresponding to a $ \mu \mathrm{s}$ switching timescale (see Appendix \ref{app:E}), the particle content is only suppressed by six orders of magnitude, which would still correspond to a signal larger than the linewidth. For standard array parameters, this is the relevant experimental regime, as we detail in Appendix.~\ref{timesec}. We find that the frequency shift in this slow-switching regime follows a similar power law scaling at short times to the full model of the atom array's response function (see inset to Fig.\ \ref{fig:suppression}(c)).

\section{Discussion} \label{Disc1}
Let us begin by examining the historical context of the dynamical Casimir effect (DCE), first proposed by Moore as particle production resulting from a moving mirror at the boundary of a one-dimensional photonic cavity \cite{MooreGeraldT.1970QTot}. Revisiting this original setting allows us to later compare modern experiments—and our proposal—to early theoretical considerations. In Moore’s work, the effect arises from a time-dependent boundary condition: the photonic cavity length changing from $r$ to $R$ due to the motion of one mirror. The resulting particle creation, particularly in the non-adiabatic regime, is closely related to that studied in Ref.~\cite{BrownEricG.2015Wdim}. However, Moore’s scheme required the mirror to move at relativistic speeds for particle creation to reach detectable levels. At lower velocities, the effect is strongly suppressed, placing such experiments well beyond practical reach. This limitation was later addressed by exploiting parametric resonance, which enables a resonant enhancement of particle production \cite{DODONOV1992309,PhysRevA.53.2664,PhysRevLett.77.615}. Such a mechanism forms the basis of the first breakthrough experimental observations of the DCE, achieved using time-dependent electric fields as boundary conditions rather than physically moving mirrors \cite{WILSONC.M2011Ootd,LahteenmakiPasi2013DCei}. Parametric coupling of fixed cavity modes also underlies recent proposals that involve mechanical acceleration of mechanical membranes \cite{WangHui2021Capp}, or accelaration radiation from vibrating atoms \cite{Dolan_2020} or similiar resonant amplification processes arising from internal system dynamics \cite{MacriVincenzo2018NDCE,PhysRevResearch.5.013221,PhysRevResearch.5.043274,PhysRevResearch.6.023320,10.21468/SciPostPhys.18.2.067}, as well as the generation of photons from a blue-detuned pulse in optomechanical set-ups \cite{RevModPhys.86.1391}. 

The parametrically amplified DCE experiments and proposals discussed above therefore could not probe particle creation from non-perturbative boundary condition changes across distinct spatial regions, as proposed in Ref.~\cite{BrownEricG.2015Wdim}, or originally analysed by Moore \cite{MooreGeraldT.1970QTot}. Instead, all proposals and observations to date have relied on parametric coupling between pairs of cavity modes induced by the mirror’s motion, with boundary conditions modifying the photonic mode only perturbatively. Consequently, the photonic mode profiles remain effectively unchanged and the dynamics reduce to an effective parametric interaction between unperturbed cavity modes.

The parametric generation of photons from the mechanical motion of a mirror boundary of a photonic cavity is a procedure commonly employed in quantum optomechanics \cite{RevModPhys.86.1391}, and we outline the connection further in Appendix~\ref{DCEcomparison}. In contrast, the particle creation described in Ref.~\cite{BrownEricG.2015Wdim}, and originally considered in Ref.~\cite{MooreGeraldT.1970QTot} is fundamentally distinct: it involves detecting photons arising from entangled, spatially distinct regions of the electromagnetic vacuum, as the field re-adjust to non-perturbative changes in the boundary conditions. This process, which underpins the particle creation phenomena considered by Moore \cite{MooreGeraldT.1970QTot} arises due to a mismatch between the local and global modes of a quantum field integrated across a non-perturbative change in boundary condition for the field modes, which is the same feature that is behind effects such as Hawking and Unruh radiation. In the present work, this feature is present in both the fast and slow switching regime - the total particle creation integrated across the switch arises due to a highly non-perturbative boundary condition change affecting the cavity mode profiles and involves the generation of entangled pairs of particles across the two spatially localised sub-cavities.
\newline 
In this way, the experimental implementation of our proposal will unlock the first experimental demonstration of the particle creation due to fundamentally split photonic mode structures as was originally considered in the original dynamical Casimir effect \cite{MooreGeraldT.1970QTot}.

\onecolumngrid
\appendix
\section{Definitions and Conventions}
\begin{equation}\label{Xkr}
\begin{split}
        \hat{X}[\mathbf{r}] &= \int d^3 \mathbf{k} \frac{e^{i \mathbf{k} . \mathbf{r}}}{(2 \pi)^{3 / 2}}\hat{X}[\mathbf{k}]\\ 
        \hat{X}[\mathbf{k}] &= \int d^3 \mathbf{r} \frac{e^{-i \mathbf{k} . \mathbf{r}}}{(2 \pi)^{3 / 2}}\hat{X}[\mathbf{r}]\\
        \delta_{i j}^{T}(\mathbf{r})&=\frac{2}{3} \delta_{i j} \delta(\mathbf{r})-\frac{1}{4 \pi r^3}\left(\delta_{i j} - \frac{3 r_i r_j}{r^2}\right)\\
       \Omega_\mathbf{k} &-\;\; \mathrm{EIT \; \; probe\; \; beam \;\; coupling \;\; strength} \\
       \Omega_p &-\;\; \mathrm{EIT \; \; control\; \; beam \;\;coupling \;\; strength} \\
       g &- \mathrm{Control} \; \mathrm{Atom} \; \mathrm{Rabi} \; \mathrm{Coupling} \; \mathrm{Strength}
\end{split}
\end{equation}
\textit{The EIT (Electromagnetically-induced-transparency) probe beam coupling strength} is a semi-classical dipole coupling between the $\ket{e} \rightarrow \ket{g}$ dipole transition for the $\mathrm{Rb}^{87}$ atom with $\ket{g} =\left|5 \mathrm{~S}_{1 / 2}, F=2, m_F=-2\right\rangle$ as the ground state and $\ket{e} = |5 \mathrm{P}_{3 / 2}, F=3, m_F=-3\rangle$ as the excited state, with transition frequency $\omega_e/2\pi = 400 \; \mathrm{THz}$. This transition has the collectively enhanced dipole moment $\Omega_\mathbf{k} = \sqrt{N}\mu^{i} E_q$, where the $\mu^{i}$ dipole moment for each individual atom quantifies the coupling to the vacuum amplitude for the incident quantum field $\sqrt{\frac{\hbar \omega}{2L A \epsilon_0}}$ (collective coupling to N atoms), where $L$ is the length scale for the longitudinally polarised electric field mode and $A$ is the transverse area of the mode. In the semi-classical limit $E_q$ is replaced with the classical field amplitude. 

\textit{The EIT control beam coupling strength} describes the dipole-moment-induced Rabi coupling strength between the weak classical field connecting the $\ket{e} = |5\mathrm{P}_{3/2}, F=3, m_F=-3\rangle$ state and the Rydberg state $\ket{r_P} = |44\mathrm{P}_{3/2}, m_J=3/2\rangle$ of a $\mathrm{Rb}^{87}$ atom. These states form the three-level EIT system used throughout this work.

The control atom is an additional $\mathrm{Rb}^{87}$ atom with energy levels that form the two control states as $\left|g^{\prime}\right\rangle=\left|5 \mathrm{~S}_{1 / 2}, F=1, m_F=-1\right\rangle$ (labelled $\ket{T}$ for simplicity through-out the text), $\left|r_P\right\rangle=\left|44 \mathrm{P}_{3 / 2}, m_J=3 / 2\right\rangle$ (labelled $\ket{R}$ through-out the text) that has a transition at $\frac{\nu}{2\pi} = 1000 \; \mathrm{THz}$. The control atom is driven between these two states with Rabi coupling strength $g$ (denoted through-out the text).

The Fourier transformation conventions in Eq.~\eqref{Xkr} simplify in the expected manner to the case of a 1-dimensional Fourier transformation, which is used in the case of time-frequency transformations.

\section{Quantum-controlled atom array response from the microscopic interactions}\label{arrayresponse}
Here, we derive the permittivity of the atomic array, and therefore its reflectivity conditioned on the state of the control atom. This will enable us to compute the array's response time, and eventually derive the quantum-controlled interaction Hamiltonian which forms the core model for this work, Eq.~\eqref{hallterms} in the main text. This builds from the theory of Refs.~\cite{Bekenstein2020,ShahmoonEphraim2017CRiL}, but goes beyond these works by deriving the dynamics and reflectivity of the atomic array conditioned on the quantum state of the control atom, and simulating the time-domain dynamics given the specified initialisation in an EIT dark state of the atomic array and extends the analysis to the interaction with weak quantum fields explicitly.

We will compare to prior transient-EIT analyses that considered evolution from ground-state preparation \cite{LiFang2025Aamf}. However, here we will solve a distinct initial-value problem: the time-domain evolution of array reflectivity starting from an already prepared EIT dark state and following a sudden interaction-induced Rydberg shift \cite{LiFang2025Aamf}. 

\subsection{Response function from the microscopic interactions for semi-classical fields}\label{response1234}

The permittivity $\alpha_\mathbf{k}$ of the atom array for a collective mode of momentum $\mathbf{k}$ is derived in Ref.~\cite{Bekenstein2020} from $\hat{H}_\mathrm{array} + \hat{U}_\mathrm{ryd}$ to be proportional to the coherence (the off-diagonal elements of the collective mode $\rho_{e g, \mathbf{k}}$) as $\alpha_\mathbf{k} = \frac{|\mu|^2}{\hbar} \frac{\rho_{e g, \mathbf{k}}}{\Omega_{\mathbf{k}}}$, where $|\mu|$ is the truncated dipole moment of the array atoms, and $\Omega_\mathbf{k}$ is the coupling strength to the incident field, and simplifying with the spontaneous emission rate, $\gamma=\frac{|\mu|^2 8 \pi^2}{3 \lambda^3 \epsilon_0 \hbar}$ (where $\lambda$ is the wavelength of resonance of the array atoms on the $\ket{g} \rightarrow \ket{e}$ transition), such that the permittivity becomes $\alpha_\mathbf{k} = \frac{3\epsilon_0\gamma \lambda^3}{8\pi^2} \frac{\rho_{e g, \mathbf{k}}}{\Omega_{\mathbf{k}}}$, with the vacuum permittivity $\epsilon_0$. The permittivity is related to the reflection coefficient $S_\mathbf{k}$  through \cite{PhysRevLett.118.113601}, $S_\mathbf{k}=i \pi\left(\frac{\lambda}{a}\right)^2 \frac{\alpha_\mathbf{k}}{\epsilon_0 \lambda^3}$, where $a$ is the spacing between atoms in the array. Now, using the co-operative corrections to the linewidth as $\Gamma_\mathbf{k} + \gamma = \gamma \frac{3}{4 \pi}\left(\frac{\lambda}{a}\right)^2$ \cite{PhysRevLett.118.113601}, this simplifies to be $S_\mathbf{k} = i \frac{(\Gamma_\mathbf{k}+\gamma)}{2} \frac{\rho_{e g, \mathbf{k}}}{\Omega_{\mathbf{k}}}$. We henceforth drop the subscript $\mathbf{k}$, leaving reference to a specific collective mode of the array of momentum $\mathbf{k}$ implicit. This directly quantifies the reflection coefficient for incident light:
\begin{equation}\label{Sikz}
    \mathbf{E}=\left[e^{i k z}+S e^{i k|z|}\right] \mathbf{E}_0,
\end{equation}
where $S$ quantifies the reflection coefficient for the incident electromagnetic field $\mathbf{E}_0$. This establishes a Dirichlet boundary condition for the electromagnetic field at $z = 0$ when $S = - 1$, as the field vanishes in the region $z > 0$ (while the expressions we give are for the right-moving fields, analogous relations hold for the left-moving ones, incident from the opposite side of the boundary at $z = 0$). In this way, we can directly relate the coherence to a reflection coefficient for the electromagnetic field, thereby quantifying the response function and reflectivity of the field through the coherence \cite{Bekenstein2020}.

The total Hamiltonian of the array-control atom system can be expressed as
\begin{equation}\label{Harray}
    \hat{H} = \hat{H}_\mathrm{array} + \hbar\omega_0 \hat{\sigma}^c_z + \hat{U}_\mathrm{ryd},
\end{equation}
where $\hat{H}_\mathrm{array}$ is the Hamiltonian of the internal states of the array atoms. The Hamiltonian of the array atoms with the EIT control beam on is 
\begin{equation}\label{Harray1}
    \hat{H}_{\text {array }}=-\Delta|r\rangle\langle r|-\Omega_{\mathbf{k}}(|e\rangle\langle g|+|g\rangle\langle e|)-\Omega_p(|e\rangle\langle r|+|r\rangle\langle e|),    
\end{equation}
where $\Delta$ is the single photon detuning. Here, starting from the total Hamiltonian in Eq.~\eqref{Harray}, we can define a permittivity conditioned on the control being in the transmissive or reflective state, there-by deriving the the action of such a quantum-controlled atomic array on the cavity modes as a quantum-controlled potential-well, applied in Sec.\ \ref{beam1}. In order to see this, we use the Rydberg-Rydberg interaction Hamiltonian, between the $i$th atom in the array and the control atom as
\begin{equation}
   \hat{V}_i = V_{ic} \ket{R}_i\bra{R}_i \otimes \ket{R}_c\bra{R}_c,
\end{equation}
with $\hat{U}_\mathrm{ryd} = \sum_i \hat{V}_i$. If we multiply the remainder of the Hamiltonian in Eq.~\eqref{Harray1}, by the identity matrix on the control atom's Hilbert space, we can thus re-arrange the Hamiltonian modeling the microscopic interaction between the control atom and each atom in the array 
as
\begin{equation}\label{Harray}
   \hat{H} = \left(\hat{H}_{\text {array }}+\hbar \omega_0 \hat{\sigma}^c_z + \sum_i V_{i c}|R\rangle_i\left\langle\left. R\right|_i\right.\right) \otimes|R\rangle_c\left\langle\left. R\right|_c\right. + \left(\hat{H}_{\text {array }}+\hbar \omega_0 \hat{\sigma}^c_z \right) \otimes|T\rangle_c\left\langle\left. T\right|_c\right.,
\end{equation}
and re-derive the equations of motion for the amplitudes of the array atoms — obtaining one set of the equations used in Ref.~\cite{Bekenstein2020} for each of the control atom’s orthogonal quantum states, $\ket{R}$ and $\ket{T}$. Denoting $c_g^j$, $c_{e,\vec{k}}^j$ and $c_{r, \vec{k}}^j$ as the probability amplitude of the array atoms to be in the $\ket{g}$, $\ket{e}$ and $\ket{r}$ states respectively, conditioned on the control atom's quantum state $j \in \{T,R\}$, we have
\begin{equation} \label{eom111}
\left[\begin{array}{l}
\dot{c}^j_g \\
\dot{c}^j_r \\
\dot{c}^j_e
\end{array}\right]=\left[\begin{array}{lll}
0 & 0 & i\Omega^*_\mathbf{k} \\
0 & -\Gamma_r & i\Omega^*_p \\
i\Omega_\mathbf{k} & i\Omega_p & -\Gamma_e
\end{array}\right]\left[\begin{array}{l}
c^j_g \\
c^j_r \\
c^j_e
\end{array}\right],
\end{equation}
 with $V^T = 0$, and  $\Gamma_r = \left[\left(\gamma_r / 2-i\left(\delta_r+V^j\right)\right)\right]$, and $\Gamma_e = \left[\left(\gamma+\Gamma_{\mathbf{k}}\right) / 2-i\left(\delta-\Delta_{\mathbf{k}}\right)\right]$. In Ref.~\cite{Bekenstein2020}, these equations of motion are solved to compute the coherence $\rho_{e g, \mathbf{k}}=c_g{ }^* c_{e, \mathbf{k}}$ in the steady state, through $\alpha(\mathbf{k})=\frac{|\mu|^2}{\hbar} \frac{\rho_{e g, \mathbf{k}}}{\Omega_{\mathbf{k}}}$ as above, and therefore the reflectivity is:
\begin{equation} \label{Sjj}   
S^j(\omega) = \frac{i(\gamma+\Gamma_\mathbf{k})\left(\delta_r+V^j\right)}{-i\left(\delta_r+V^j\right)(\gamma+\Gamma_\mathbf{k} - 2i(\delta-\Delta_\mathbf{k}))+2\left|\Omega_{\mathrm{p}}\right|^2},
\end{equation}
where here $\delta=\omega_{e g}-\omega$ is the detuning of the light from the atomic resonance, and $\gamma$ denotes the radiative lifetime of the state $\ket{e}$. It is standard now to apply this control-dependent reflectivity derived for weak incident classical fields, carries over to weak incident quantum fields in the absence of non-linearities. This quantum-mechanical derivation of the array coherence accounts for the Rydberg interaction energy $\hat{U}_\mathrm{ryd}$, which shifts the atoms’ energy levels and thereby modifies the coherence - and hence the effective permittivity - of the atomic array, tuning it from highly reflective to transparent. Finally, we briefly comment that this model of the collective mode as if it were one giant atom is fairly standard in the weak excitation limit of the collective mode, as was used in Ref.~\cite{Bekenstein2020}.

\subsection{Time-domain Switching dynamics}\label{timesec}
\subsubsection{Initial Conditions}
In this section, we now solve for the EIT time-domain dynamics given initialisation of the collective mode in an EIT Dark state. Here, $\Delta$ and $\Gamma$ are the co-operative corrections to $\delta$ and $\gamma$ arising from dipole-dipole interactions, and $\delta_r = \omega_{r g}-\omega_{\mathrm{p}}-\omega$, is the two-photon detuning.   In order to model the time-domain dynamics, we consider that the Hamiltonian in Eq.~\eqref{Harray1} has the EIT dark state as an eigenstate:
\begin{equation}
\left|D\right\rangle=\cos \theta|g\rangle-\sin \theta|r\rangle,
\end{equation}
with $\tan(\theta) = \frac{\Omega_\mathbf{k}}{\Omega_p}$, for small $\theta$, this dark state becomes
\begin{equation}\label{Dgr}
    \left|D\right\rangle \simeq |g\rangle-\frac{\Omega_\mathbf{k}}{\Omega_p}|r\rangle.
\end{equation}
With the quantum-controlled EIT model established from the previous section, we can now undertake detailed modeling for the various regimes of the array's response function. In the main text, there are four regimes discussed for the scaling of the switching of the array's reflectivity which are presented as the core results. These different regimes can be accessed depending on the initial state of the array, as well as the strength of the Rydberg interaction as compared to the EIT linewidth $\Gamma_\mathrm{EIT} = \frac{\Omega_p^2}{\Gamma_r}$.

Next, we solve for the time-domain dynamics of the reflectivity given in Eq.~\eqref{eom111}. We first check that in the absence of a Rydberg shift, and initializing the collective mode in the ground state, that we qualitatively re-obtain similiar dynamics that was recently derived in Ref.~\cite{LiFang2025Aamf} for the transmissivity (the imaginary component of the susceptibility) without extending to the full four-level model of Ref.~\cite{LiFang2025Aamf}. Next, rather than initialising in the ground state as was considered for a different four-level EIT system model in Ref.~\cite{LiFang2025Aamf}, we compute the rise of the coherence and therefore  of the reflectivity out of the initial EIT dark state after a switch-on of the Rydberg interaction (resulting in the energy level shift of the array atoms). We find that for typical array parameters (as discussed in Section~\ref{expconsider1}), the rise of the coherence, and therefore the susceptibility (and therefore the transmission and reflection coefficients),  change on the time-scale given by the linewidth of the array atoms, as simulated in Fig.~\ref{ref} (a).

\subsubsection{Regime I}

The first regime discussed in the main text—corresponding to linewidth-limited slow switching—is obtained by solving the system of ODEs in Eq. \eqref{eom111} under the assumption that the system is initially prepared in the dark state. The time-domain solution is solved directly from the ODEs, which describes the transition from the dark state to the reflective state of the atomic array. For the standard array parameters considered in Section \ref{expconsider1}, where the Rydberg interaction strength is much larger than the EIT linewidth ($V^R \gg \Gamma_{\mathrm{EIT}}$), the susceptibility switches on over nanosecond timescales. This places the system firmly in the slow-switching regime, in which the field modes adjust adiabatically to the changing boundary conditions with deviations from the vacuum state arising only through non-adiabatic corrections to the adiabatic theorem.

Although the switching is slow compared to the optical transition frequency $\omega_e \sim 400 \; \mathrm{THz}$, at early times the reflectivity can be still considered to slam to a low reflectivity, that generates measurable photon content and an associated frequency shift. The total effect can be obtained by integrating these non-adiabatic corrections over the full switching interval, as computed in Section \ref{multinon}. This process corresponds to vacuum particle creation driven by highly non-perturbative changes in the boundary conditions forcing the vacuum to reorganize into split cavity modes - a significant step beyond previous dynamical Casimir experiments and in line with the original effect proposed by Moore \cite{MooreGeraldT.1970QTot}.

Accordingly, for these standard array parameters we model the reflectivity as $r(t - t_0) = 1 - \exp\!\big[-\kappa \Gamma_e (t - t_0)\big]$, where $t_0$ is the switch-on time and $\kappa$ is extracted from the full ODE solution shown in Fig. \ref{ref}. This form captures a switching time on the order of the linewidth and matches the slow-switching toy model, particularly the short-time power-law scaling of the reflectivity. We emphasize that these parameters allow for particle creation through non-adiabatic corrections to the adiabatic theorem with standard array parameters ($\mathrm{MHz}$ array atom linewidths combined with $\mathrm{MHz}$ EIT Control beam coupling strengths), in the regime of highly non-perturbative boundary condition changes. Furthermore, increasing the atomic linewidth by several orders of magnitude through extreme Purcell enhancement would drive the system into the fast-switching regime, enabling access to the ideal frequency shift for rapid switching.

\subsubsection{Regime II}

\begin{figure*}[h]
\begin{center}
\includegraphics[width=\linewidth]{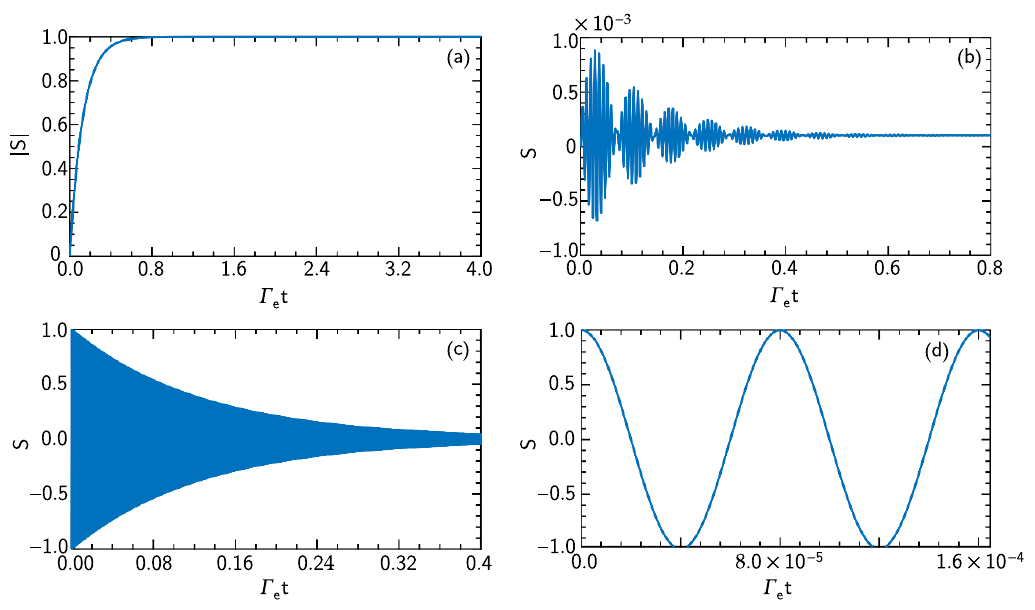}
    \caption{\label{ref} (a) Time-dependence of the reflectivity of the atomic array initialised in the EIT dark state, after a speed of light switch on of the Rydberg potential energy $V^R$. The time is renormalised to the linewidth, with $\Omega_p = 10\Gamma_e$, $V^R = 10^5\Gamma_e$, $\Gamma_e = \Gamma_r$. (b)~Time-dependence of the reflectivity of the atomic array initialised in the EIT dark state, after a speed of light switch on of the Rydberg potential energy $V^R$. The time is renormalised to the linewidth, with $\Omega_p = 10^4\Gamma_e$, $V^R = 10^3\Gamma_e$, $\Gamma_e = \Gamma_r$. In this regime, the Rydberg interaction is not strong enough to switch the array to near unit reflectivity. However, fast oscillations are observed before reaching the steady state, on a time-scale proportional to $\Omega_p$. (c) Time-dependence of the reflectivity after initialising in the highly reflective state with the Rydberg shift on in the regime $V^R \gg \frac{\Omega_p^2}{\Gamma_r}$, and a speed of light switch-off of the Rydberg interaction energy, in this regime for which $\Omega_p = 10^4\Gamma_e \gg  \Gamma_e$, fast oscillations are observed on the time-scale $\Omega_p$, with a decay envelope on the time-scale of the linewidth. (d) Inset into the fast oscillations displayed in (c). Code for the simulations is attached as an ancillary file. Scaling the EIT control beam coupling strength all the way down to $\mathrm{MHz}$ coupling strengths yields the same qualitative behaviour in (b) and (c) with slower oscillation speeds, this enables parametric particle production in this regime going from $\mathrm{MHz} - \mathrm{GHz}$ scale oscillation speeds. }
\end{center}
\end{figure*}

The second regime discussed in the main text corresponds to initialising in the EIT Dark state, and examining the dynamics for $V^R \ll \Gamma_\mathrm{EIT}$. The system exhibits oscillations of low reflectivities on a time scale proportional to the EIT probe beam strength $\propto \Omega_p$. For EIT probe beams such that $\Omega_p/2\pi \sim \;\mathrm{MHz} - 10 \; \mathrm{GHz}$, such switching to low reflectivities as depicted in Fig.~\ref{ref} (b) is realistic, enabling parametric vacuum particle content production to be observed (as for such low reflectivities this corresponds to a parametric particle creation effect, albeit with perturbative boundary condition changes - as the system slams to a low reflectivity only parametrically coupling previously unperturbed cavity modes as we model in Section~\ref{beam1}).

\subsubsection{Regime III}

The final regime involves using strong EIT control beams on the order of $\Omega_p/2\pi \sim 100 \; \mathrm{GHz}$, initiating in a highly reflective state, and we simulate the array's susceptibility to flip to rapidly oscillate about zero reflectivity on time-scales proportional to $\Omega_p$. In order to access this regime, it is imperative that the EIT control beam reach this $100 \; \mathrm{GHz}$ strength - albeit this is a very experimentally challenging regime for a driving strength on the $\ket{e} \rightarrow \ket{r}$ transition for the array atoms. 

\subsection{Response function from the microscopic interactions for quantum fields}\label{qfield1}
In this section, we aim to explicitly re-derive the result from the previous section for the control-dependent response to incident weak classical fields explicitly now for weak quantum fields (Extending the analysis of Refs.~\cite{Bekenstein2020,ShahmoonEphraim2017CRiL} in which this was derived for weak classical fields). We now derive this explicitly starting from the full QED Hamiltonian for non-relativistic matter interacting with light - prior to gauge-fixing, 
\begin{equation}\label{nrr11}
    H=\frac{1}{2} m \dot{\mathbf{r}}^2+\frac{1}{2} \int d^3 x\left(\mathbf{\tilde{\Pi}}^2+\mathbf{B}^2\right),
\end{equation}
where $\mathbf{\tilde{\Pi}} = \dot{A}$ and $\mathbf{B} = \nabla \times \mathbf{A}$ are the electric and magnetic fields with classical Poison bracket,
\begin{equation}
\begin{split}
        \left\{A_{i}(\boldsymbol{r}), \tilde{\Pi}_{k}\left(\boldsymbol{r}^{\prime}\right)\right\}&=\delta_{i k} \delta^{(3)}\left(\boldsymbol{r}-\boldsymbol{r}^{\prime}\right),\\
        \left\{r_{i}, p_{k}\right\} &= \delta_{ik},
\end{split}
\end{equation}
computed through functional derivatives, while $m\dot{\mathbf{r}} =\mathbf{p} - q\mathbf{A}$ follows from the Newton-Lorenz Law. If we further restrict to the Couloumb gauge, this simplifies the Poisson brackets to be,
\begin{equation}
\begin{split}
        \left\{A_{m}(\boldsymbol{r}), \tilde{\Pi}_{n}\left(\boldsymbol{r}^{\prime}\right)\right\}&=\delta_{mn} \delta^{T}\left(\boldsymbol{r}-\boldsymbol{r}^{\prime}\right),\\
        \left\{r_{i}, p_{k}\right\} &= \delta_{ik},
\end{split}
\end{equation}
where $m,n \in \{1, 2\}$ run over the two independent components of the remaining two degrees of freedom in the Coloumb gauge, while $i,k \in \{1,2,3\}$ run over three-independent components of the canonical position and momentum of the matter.  If we now canonically quantise promoting the Poisson brackets to commutators,
\begin{equation}
\begin{split}
    \left[\hat{r}_i, \hat{p}_k\right] &= i\delta_{ik},\\
    \left[\hat{A}_m(\boldsymbol{r}), \hat{\tilde{\Pi}}_n\left(\boldsymbol{r}^{\prime}\right)\right]&=i\delta_{mn} \delta^{T}\left(\boldsymbol{r}-\boldsymbol{r}^{\prime}\right).
\end{split}
\end{equation}
We can now simplify the dynamics by applying a Power-Zienau-Woolley transformation \cite{Stokes_2019}
\begin{equation}
    R_{\alpha \alpha^{\prime}}:=\exp \left[\mathrm{i} \int d^3 \mathbf{x}\left[\alpha^{\prime}\mathbf{P}_{\mathrm{T}}(\mathbf{x})-\alpha\mathbf{P}_{\mathrm{T}}(\mathbf{x})\right] \cdot \mathbf{A}_{\mathrm{T}}(\mathbf{x})\right],    
\end{equation}
now when this transformation is applied to the full Hamiltonian of Eq.~\eqref{nrr11}, produces the Hamiltonian (transforming from the Coloumb gauge $\alpha = 0$ to the $\alpha$ gauge), through $R_{\alpha, 0}\hat{H}_{\mathrm{Coloumb}}R^\dagger_{\alpha, 0}$. In this case the Coloumb gauge Hamiltonian transforms to 
\begin{equation}
    H=\frac{1}{2 m}\left[\mathbf{p} - q(1-\alpha) \mathbf{A}_{\mathrm{T}}(\mathbf{0})\right]^2+V(\mathbf{r})+\frac{1}{2} \int d^3 x\left(\left[\mathbf{\tilde{\Pi}}(\mathbf{x}) + \alpha d_j \delta_{i j}^{\mathrm{T}}(\mathbf{x})\right]^2+\left[\nabla \times \mathbf{A}_{\mathrm{T}}(\mathbf{x})\right]^2\right),
\end{equation}
using the polarisation operator as $\hat{\mathbf{P}}_{T} = \mu_j \delta_{i j}^{\mathrm{T}}(\mathbf{x})$, with $\mu_j = qr_j$,  we can now collect, 
\begin{equation}\label{Hmca}
    \begin{aligned}
    H_{\mathrm{m}}~=~& \frac{\mathbf{p}^2}{2 m}+V(\mathbf{r}), \\
    H_{\mathrm{c}}~=~&\epsilon_0 \int \mathrm{d}^3 \mathbf{x}\left[ \frac{1}{2}\left(\boldsymbol{E}(\mathbf{x})^2+(\curl \mathbf{A})^2\right) + \alpha \hat{\mathbf{P}} \cdot \boldsymbol{E}(\mathbf{x}) +\frac{\alpha^2}{2} \hat{\mathbf{P}}^2\right] ,\\
    V^\alpha= & -\frac{q}{m}(1-\alpha) \mathbf{p} \cdot \mathbf{A}_T(0)+\frac{e^2}{2 m}(1-\alpha)^2 \mathbf{A}_T(0)^2,
\end{aligned}   
\end{equation}
where $\mathbf{E} = -\frac{\mathbf{\tilde{\Pi}}}{\epsilon_0}$ is the Coloumb-gauge Electric field.

Now if we fix $\alpha = 1$ to the multi-polar gauge and neglect second order terms in the dipole operator (weak excitation limit that we will use through-out this work) - the Hamiltonian reduces to the standard Jaynes-Cummings model with the dipole interaction term: 
\begin{equation}\label{Hint1111}
    \hat{H}_{\mathrm{int}} = \epsilon_0\int\mathrm{d}^3\mathbf{x} \; \hat{\mathbf{E}}_T(\mathbf{x}) \cdot \hat{\mathbf{P}}_T(\mathbf{x}),
\end{equation}
between quantum field modes and the atomic array. In the Heisenberg picture, a time-dependent dipole coupling term re-normalises the canonical pairs for the photonic subsystem $\boldsymbol{\Pi}_\alpha= \hat{\mathbf{P}}_T + \mathbf{\tilde{\Pi}}$. We henceforth, truncate to the lowest two energy levels of the polarisation operator, and assume under the rotating wave-approximation that the interaction hamiltonian simplifies to be $\hat{H}_{\mathrm{int}} = \epsilon_0\int \mathrm{d}^3 \mathbf{x}\left(\hat{\mathbf{E}}^{+}_T(\mathbf{x}) \cdot \hat{\mathbf{P}}^{-}_T(\mathbf{x}) + \hat{\mathbf{E}}^{-}_T(\mathbf{x}) \cdot \hat{\mathbf{P}}^{+}_T(\mathbf{x})\right)$, (and at select points through-out the text we restore the counter-rotating terms), where the super-script denotes the positive and negative frequency components of the field respectively. We henceforth use the expected generalisation from a single dipole to a sum over $N$ atomic dipoles in Eq.~\eqref{Hmca}, producing a self-energy polarisation operator term $\sum_{i,j} \hat{\mathbf{P}}_i\hat{\mathbf{P}}_j$ that will produce an excited state energy shift once we convert to the collective mode.

The positive frequency component of the total polarization operator (for all dipoles that couple to the electromagnetic field) for the individual dipole sources can be written as a sum over all of the point-like atomic sources that make up the array,
\begin{equation}        \widehat{\mathbf{P}}^{+}_T(\mathbf{x}, t) \simeq \sum_l \hat{d}^l(t) \delta^l\left(\mathbf{x}-\mathbf{x}_l\right),   \end{equation}
where $\hat{d}^l(t)=\mu_{\mathrm{ge}}^l \hat{\sigma}_{\mathrm{ge}}^l(t)$, is the dipole operator of the l-th atom and $\mu^{l}_{\mathrm{ge}}$ its dipole matrix element. In this formulation, for the resulting wave equation, the atomic array appears as a delta-function barrier in the equations of motion for the field - a representation that we will examine further in Section \eqref{beam1}. In Section \eqref{beam1}, we will also use that Coloumb gauge photonic field operators and note that the bare Coloumb gauge free field Hamiltonian $\epsilon_0\int \mathrm{d}^3 \mathbf{x}\left[\frac{1}{2}\left(\tilde{\boldsymbol{\Pi}}(\mathbf{x})^2+(\boldsymbol{\nabla} \times \mathbf{A})^2\right)\right.$, after restriction to 1-D by integrating out the transverse co-ordinates spatial co-ordinates $\mathbf{y}, \mathbf{z}$, this can be mapped 1-1 to the free Hamiltonian of a scalar field with the identification $A_y \rightarrow \frac{\phi}{\sqrt{\epsilon_0}}$ and $E_y \rightarrow \frac{\pi}{\sqrt{\epsilon_0}}$ (with longitudinal polarisation in the $x - \mathrm{direction}$) - a simplification that we explore further in Section~\ref{beam1}, i.e. that it follows that the standard expansion of the scalar field in terms of cavity mode functions $\hat{\phi}(x) = \sum_{m=1}^{\infty}\left(f_m(x) \hat{a}_m+f_m^*(x) \hat{a}_m^{\dagger}\right)$ from Ref.~\cite{BrownEricG.2015Wdim} follows - where the solutions are the mode functions given in the main text - satisfying the global cavity Dirichtlet boundary conditions $\phi(0) = \phi(L) = 0$. Henceforth we restrict to $\epsilon_0 = 1$.

Next, we extend the Heisenberg-Langevin equations for the atomic array, to a three-level system and converting to be in terms of the collective mode operators $\sqrt{N}\hat{\sigma}_{\mathbf{k}}^{+}=\sum_i \hat{\sigma}_i^{+}$, we obtain the dynamics of the polarisation operator from the interaction Hamiltonian in Eq.~\eqref{Hint1111} (in a rotating frame at the cavity frequency with atom-cavity detuning $\Delta_e$, i.e. we convert to the rotating frame of $\hat{U}(t) = \exp(+i\omega_c\hat{\sigma}^{+}\hat{\sigma}^-)$): 
\begin{equation}\label{HL1}     \dot{\hat{\sigma}}^{-,\mathbf{k}}_{R} = -i\sqrt{N} {\mu}^{\mathbf{k}}\hat{E}(r,t)(\hat{\sigma}^\mathbf{k}_\mathrm{ee} - \hat{\sigma}^\mathbf{k}_\mathrm{gg})\mathrm \; + \left(i \Delta_{e} - \Gamma_e\right) \hat{\sigma}_{R}^{-, \mathbf{k}} + \sqrt{2\Gamma_e} \hat{P}_{\mathrm{i} n}(t),  
\end{equation}
where we have assumed a loss-less field mode of the electromagnetic field, where-as in the semi-classical calculation we define control-conditioned dynamics with the label $j \in \{T,R\}$. Promoting this state-dependent quantity to an operator that quantifies the reflected electric field conditioned on the quantum state of the control. The input noise operator satisfies the following correlation function relations (for negligible thermal excitation):
\begin{equation}
\hat{S}_{\mathrm{in}}(t)=\sqrt{2\Gamma_e}\,\hat{P}_{\mathrm{in}}(t),\qquad
\langle \hat{S}_{\mathrm{in}}(t)\rangle = 0,\qquad
\langle \hat{P}_{\mathrm{in}}(t)\hat{P}_{\mathrm{in}}^{\dagger}(t')\rangle \approx \delta(t-t'),
\end{equation}
where we have used that in the weak excitation limit, we can use the correlators that are typically used for bosonic mode operators. Equation~(A19) corresponds to the simplified Heisenberg--Langevin equations under the following assumptions: firstly, the atoms are initialised in the EIT dark state up to $\mathcal{O}\!\left((\Omega_k/\Omega_p)^2\right)$, with $\Omega_k = \sqrt{N} g_q$, as well as that in the reflective (blockaded) branch the EIT control beam is shifted off resonance by the Rydberg interaction, so its coupling can be neglected (equivalently, in the rotating frame of the $\ket{r}$-transition the Rydberg-shifted term is counter-rotating and is dropped). Next, it can be checked that the re-including this term and initialising the atom in the EIT Dark state of Eq.~\eqref{Dgr} the coherence and therefore the reflectivity remains zero for all times, using $\hat{\sigma}_{\mathrm{ee}}-\hat{\sigma}_{\mathrm{gg}}\approx - \hat{I} $ (weak-probe limit). With these simplifications, the Heisenberg--Langevin equation in~(A19) has the Fourier-domain solution $\hat{\sigma}_{R}^- (\omega)=-i \frac{\sqrt{N} \mu^{\mathbf{k}} \mathbf{E}(\omega)}{i(\omega - \omega_e) - \frac{\Gamma_e}{2}}$, substituting this result into the reflection coefficient derived above and working in the near-resonant regime $\omega-\omega_e \ll \frac{\Gamma_e}{2}$ we recover exactly the same coefficient as in Refs.~\cite{Bekenstein2020,ShahmoonEphraim2017CRiL}. We now use that the positive frequency component of the electric field at the barrier in this frame is
\begin{equation}
\hat{E}^{+}(\mathbf{r},t)=g_q\hat{a}\exp(-i\omega_c t)\,f(\mathbf{r},t),
\end{equation}
where $f(\mathbf{r},t)$ is the standing-wave mode function. Directly integrating the initial equation for the coherence (and assuming close to resonance with the atomic transition frequency) - up to noise terms we obtain, 
\begin{equation}\label{sge}
    \hat{\sigma}_{R}^-(t)=+i \sqrt{N} \mu^{\mathbf{k}} E_q  \int_{0}^t \mathrm{~d} \tau f(\mathbf{r},\tau) \left(\hat{a}(\tau) e^{\left((t-\tau)\left(i \Delta_e - \Gamma_{\mathrm{e}}\right)\right)}\right) + \mathrm{noise \; term},  
\end{equation}
where $f(\mathbf r,t)$ is, in general, time dependent (we take $t = 0$ to be the switching time), and $\Delta_e$ is the detuning from the atomic resonance frequency in a rotating frame close to the atomic resonance. From here on, we drop the control-conditioned subscript and leave the control dependence implicit: the linewidth-limited polarisation operator refers to the reflective branch, while no barrier refers to the transmissive branch (we also leave implicit projectors on the photonic degree of freedom). Coupling to higher-order cavity modes is suppressed by their detuning set by the free spectral range (FSR), so we retain only the single cavity mode that lies within the array’s reflectivity bandwidth. Finally, assuming the mode function does not change appreciably during the switch, we may treat $f(\mathbf r,t)\approx f(\mathbf r)$ and simplify the expression by neglecting counter-rotating terms, (within the weak-coupling, Markov, and rotating-wave approximations):
\begin{equation}\label{muPt}
    \hat{\mathbf{P}}^{+}_T(t) = \frac{2 i \sqrt{N} (\mu^{\mathbf{k}})^2 \hat{E}^+(t) }{\Gamma_e}\left(1 - \exp(-\left(\Gamma_e t \right) )\right) +\sqrt{2\Gamma_e} \int_{0}^t d \tau e^{-\left(i \Delta_e + \Gamma_e\right)(t-\tau)} \hat{P}_{\mathrm{in}}(\tau),
\end{equation}
This now produces a barrier term - supplemented with an in-homogeneous noisy source term that defines a state-dependent barrier (or susceptibility) for the cavity modes (as a block-diagonal solution for the array dynamics conditioned on the quantum state of the control). Counter-rotating contributions to the susceptibility are sub-leading and therefore ignored (while such leading order contributions to the effective susceptibility still produces dynamics on the squeezing terms for the photonic cavity field modes as we show later). 

We note that the Hamiltonian used to compute the control-conditioned dynamics of the array in this section is
\begin{equation}
\begin{split}
        \hat{H} &= g_q(\hat{\sigma}^{+}\hat{a}^{\dagger}\exp(2i\omega_ct) + \hat{\sigma}^{-}\hat{a}\exp(-2i\omega_ct))  + g_q(\hat{\sigma}^{-}\hat{a}^{\dagger} + \hat{\sigma}^{+}\hat{a}))\otimes \ket{R}\bra{R} \\&+ (g_q(\hat{\sigma}^{+}\hat{a}^{\dagger}\exp(2i\omega_ct) + \hat{\sigma}^{-}\hat{a}\exp(-2i\omega_ct))  \\&+ g_q(\hat{\sigma}^{-}\hat{a}^{\dagger} + \hat{\sigma}^{+}\hat{a}) + \Omega_p(\hat{\sigma}_{re}^{+} + \hat{\sigma}_{re}^{-} ) )\otimes \ket{T}\bra{T}  \\&+g\left(|T\rangle\langle R| e^{i \omega_D t}+\text { h.c }\right) + V\ket{R}\bra{R} + (\Delta_e\sigma^+\sigma^-)\otimes(\ket{R}\bra{R} +\ket{T}\bra{T} ).
\end{split}
\end{equation}
In the transmissive branch, we include the EIT control-beam coupling on the $\ket{e}\rightarrow\ket{r}$ transition through the Hamiltonian term $\Omega_p(\hat{\sigma}_{re}^{+} + \hat{\sigma}_{re}^{-} )$ on the $\ket{e} \rightarrow \ket{r}$ transition. The Heisenberg–Langevin equations considered here are obtained from this Hamiltonian, with the additional inclusion of the usual decay term from $\ket{e} \rightarrow \ket{g}$ at the rate $\Gamma_e$ and from $\ket{r} \rightarrow \ket{e}$ at the rate $\Gamma_r$ (in both branches of the control, however as mentioned earlier, in the reflective branch, the $\ket{r} \rightarrow \ket{e}$ term is strongly counter-rotating). In the weak-drive regime for the control atom, the control-drive-induced contribution to the evolution of the control-conditioned polarization operator is sub-leading and can be neglected (in other words additional effects from this term that will henceforth be neglected are sub-leading at the order $\frac{g}{g_q}$, for weak control atom coupling strengths $g$). 

\section{Quantum-controlled Hamiltonian }\label{QCHS}
In this section, we now apply the state-dependent reflectivity of the previous section to derive the energy distribution of the field modes conditioned on the quantum-state of the control.

\subsection{QED back-action}\label{qedback1}
The Langevin equation considered in the previous section, define block-diagonal state-dependent barrier structures for the photonic field. In addition, we restrict to a single cavity mode that is significantly modified by the barrier, and we place the barrier near a node of the two sub-cavity modes (as discussed in Sec.\ \ref{beam1}). Under these assumptions, Eq.~\eqref{muPt} gives the solution for the polarisation field. With this solution for the positive frequency component of the electric field and our Coloumb-gauge-quantisation scheme, we can express the free Hamiltonian of the electromagnetic field conditioned on the control atom state (after having converted with an inverse frame transformation on the photonic mode to restore the free-energy term)
\begin{equation}\label{timed1}
    \hat{H}_R =\omega_0\hat{a}^\dagger\hat{a} + f(r,t)^2i\omega_0\hat{a}^\dagger\hat{a}\left(1-\exp \left(-\Gamma_et \right)\right) +\mathrm{counter-rotating-terms},
\end{equation}
where $f(x,t)$ is the in general value of the time-dependent mode function at the barrier. An effective Hamiltonian here is extracted for the counter-rotating terms as $\hat{H}_\mathrm{counter} = i \omega_0\left(f(r, t)^2\right)\left(1-\exp \left(-\Gamma_et \right)\right)\left(\frac{1}{2} \hat{a}^{\dagger} \hat{a}^{\dagger}-\frac{1}{2} \hat{a} \hat{a}\right)$. Using the quantisation condition in Eq.~\eqref{jumpcontinuity}, and expanding about reflectivities close to unity, and taking the large $n$ limit (where $n$ is the global cavity mode number), we find $\frac{\dot{\omega}(t)}{\omega(t)} \sim \frac{\dot{r}(t)}{n}$, where $r(t) = (1 - \exp(-\Gamma_e t))$. In essence, Eq.~\eqref{timed1} is extracted as an effective Hamtilonian when the leading order solution for the polarisation in Eq.~\eqref{muPt} is substituted into the Heisenberg-Langevin equation for the fixed cavity modes:
\begin{equation}\label{Oadag}
    \dot{O}(\hat{a},\hat{a}^\dagger )=i\left[\omega_0 \hat{a}^{\dagger} \hat{a},O(\hat{a},\hat{a}^\dagger) \right] + ig_q\left[\hat{a}^{\dagger} + \hat{a}, O(\hat{a},\hat{a}^\dagger)\right]\hat{\sigma}_{\mathrm{ge}}^{\mathbf{k},-}(t) + ig_q\left[\hat{a} + \hat{a}^{\dagger}, O(\hat{a},\hat{a}^\dagger)\right]\hat{\sigma}_{\mathrm{ge}}^{\mathbf{k},+}(t).
\end{equation}
Restricting the observable to be just the annihilation operator and converting with an inverse frame transformation on the photonic mode to restore the free evolution piece, one can infer the effective non-hermitian Hamiltonian for the photonic cavity modes in Eq.~\eqref{timed1} generating dynamics as a commutator of the non-hermitian Hamiltonian with the annihilation operator (for more general observables, the dynamics implied by Eq.~\eqref{Oadag} is more general than just the time evolution generated by the commutator with this non-hermitian Hamiltonian - we refer to Ref.~\cite{TobarGermain2026MQSa} for this dynamics - in brief, probability conserving jump terms manifest, however, if we restrict to just the dynamics of single annihilation or creation operators, the generalised dynamics reduces to the commutator with the non-hermitian hamiltonian defined in Eq.~\eqref{timed1}). Then through a re-definition of the creation and annihilation operators in terms of a time-dependent frequency, one can define effective time-dependent cavity modes - we solve for the exact dynamics of this in Appendix.~\ref{EPP}.

In summary, we started from the full light–matter Hamiltonian, including the counter-rotating photonic terms. In the weak-coupling regime, the array is adiabatically eliminated while retaining only the leading order atomic-induced susceptibility in Eq.~\eqref{muPt}. The counter-rotating contributions to the atomic-induced susceptibility give parametrically smaller corrections and are therefore neglected. The resulting leading order susceptibility is then inserted into the full photonic Heisenberg equation, where it acts on the complete field quadrature and hence generates both number-conserving and squeezing (we use the atomic array’s rotating contribution to the field-mode susceptibility, while retaining squeezing terms in the full field-mode dynamics.).

Finally, if we take the effective input-out relation that follows from the same Hamiltonian  \cite{PhysRevA.111.053712}, 
\begin{equation}
    \hat{\mathbf{E}}^{+}(\mathbf{r}, t)=\hat{\mathbf{E}}_0^{+}(\mathbf{r}, t)+ \frac{\omega_0^2}{c^2} \sum_i \mathbf{G}\left(\mathbf{r}, \mathbf{r}_i ; \omega_0\right) \cdot \mathbf{\mu}_i \hat{\sigma}^{i,-}(t),  
\end{equation}
where $\mathbf{\mu}_i$ is the dipole moment. We next convert to the collective mode of the dipole operators - and use the defining relation for the Green's function for the global cavity (continuing to work in units where $\varepsilon_0 = 1$), 
\begin{equation}
\left[\nabla_a \times  \nabla_a \times-\frac{\omega_0^2}{c^2}\right] \mathbf{G}\left(\mathbf{r}, \mathbf{r}_i, \omega\right)=\boldsymbol{\delta}\left(\mathbf{r}-\mathbf{r}_i\right),
\end{equation}
after substituting the lowering-operator solution from Eq.~\eqref{muPt} back into the electric-field expression, we obtain a model that is equivalent to the electric field interacting with a delta barrier source term from a full quantum-field theory perspective as we examine in Section~\ref{beam1}, provided we restrict to one dimension and interpret the curl operator as acting on transversely varying electric-field components with longitudinal polarization. Taken Together - and that the input mode $\hat{\mathbf{E}}_0^{+}(\mathbf{r}, t)$ satisfies the relation in the absence of sources $\left[\nabla \times \nabla \times-\frac{\omega^2}{c^2}\right] \mathbf{E}_0(\mathbf{r}, \omega)=0$, then we obtain the following equation for the positive frequency component of the EM field:
\begin{equation}
    \nabla_a \times \nabla_a \times\hat{\mathbf{E}}^{+}(\mathbf{r}, t) -\frac{\omega_0^2}{c^2}\hat{\mathbf{E}}^{+}(\mathbf{r}, t) = \sum_i \frac{\omega_0^2}{c^2}\delta\left(\mathbf{r}-\mathbf{r}_i\right)\mu_i \hat{\sigma}^{i,-}(t).
 \end{equation}
 Finally, restricting to one dimensions and assuming longitudinally polarised cavity modes, we obtain 
 \begin{equation}
        \pdv[2]{\mathbf{E}^{+}(x, t)}{x} - k^2\mathbf{E}^{+}(x, t) = k^2 \sqrt{N}\mu_\mathbf{k}\hat{\sigma}^{-,\mathbf{k}}(t)\frac{\delta\left(x-r\right)}{A}.
  \end{equation}
  where $r$ is the longitudinal location of the atomic array in the photonic cavity. This wave equation for the electric field enables us to examine the effect of the array on the spatial mode functions of the cavity, as we study in the next section. 

\subsection{Quantum-Controlled Reflectivity Mapped to Scalar Field Theory}\label{beam1}
Here we further justify the Hamiltonian for the photonic cavity conditioned on the quantum-control variable that forms Eq.~\eqref{hallterms} in the main text. We approach this by matching the previously derived reflectivity to the imperfect reflectivity of a delta barrier term. This further enables us to examine the effect of the barrier on the cavity mode functions.

We take as a starting point, the Hamiltonian density of a free scalar field $\phi(x,t)$,with the understanding as in Ref.~\cite{BrownEricG.2015Wdim} that it can be used as a model for a single polarisation of the electromagnetic field confined to an optical cavity. This can be seen explicitly by associating the electromagnetic vector potential in the Couloumb gauge to be $A_y \rightarrow  \frac{\phi}{\sqrt{\epsilon_0}} $, $E_y \rightarrow \frac{\dot{\phi}}{\sqrt{\epsilon_0}}$ - in which case the photonic QED Hamiltonian in the Coloumb gauge from Section~\ref{qfield1} reduces to
\begin{equation}\label{Hphi}    \hat{H}=\int\mathrm{d}^3x \frac{1}{2}\left[(\dot\phi(x))^2+\left(\partial_x \phi(x)\right)^2\right],
\end{equation}
and impose Dirichtlet boundary conditions $\phi(0) = \phi(L) = 0$. We supplement this free field Hamiltonian density with the quantum-controlled version of the scattering relation in Eq.~\eqref{Sikz}. Here, the effect is implemented as a delta function potential with coupling strength conditioned on the quantum state of the control: 
\begin{equation}\label{Hintg}
    \mathcal{\hat{H}}_\mathrm{int} = g \delta(x-r)\hat{\phi}_\Delta^2  \otimes |R\rangle_c\langle\left. R\right|_c +  \hat{I}\otimes |T\rangle_c\langle\left. T\right|_c.
\end{equation}
Here, $\hat{\phi}_\Delta$ denotes the component of the field within the array’s reflectivity bandwidth. In the transmissive branch, there is no coupling between light and the atomic array because the array is in an EIT dark state (Appendix~\ref{Harray}).  
We next aim to construct the full quantum-controlled Hamiltonian density. Firstly, we multiply the free Hamiltonian density in Eq.~\eqref{Hphi}, by identity on the field DoF times identity on the control Hilbert space $\hat{I} \otimes \hat{I}_c $, where $\hat{I}_c~=~|R\rangle_c\left\langle\left.R\right|_c\right.~+~|T\rangle_c\left\langle\left. T\right|_c\right.$. Modelling the presence of a mirror as a delta function potential for the field, conditioned on the state of the quantum-control as in Eq.~\eqref{Hintg}, allows us to construct the following Hamiltonian density for the quantum field conditioned on the state of the quantum control system:
\begin{equation}\label{Hrr1}
    \hat{\mathcal{H}}=\left(\frac{1}{2}\left[\left(\frac{\mathrm{d}}{\mathrm{d}t}\hat{{\phi}}\right)^2+\left(\partial_x \hat{\phi}\right)^2\right] + g \delta(x-r) \hat{\phi}_\Delta^2 \right)\otimes|R\rangle_c\left\langle\left. R\right|_c\right. + \left(\frac{1}{2}\left[\left(\frac{\mathrm{d}}{\mathrm{d}t}\hat{{\phi}}\right)^2+\left(\partial_x \hat{\phi}\right)^2\right]  \right)\otimes|T\rangle_c\left\langle\left. T\right|_c\right..
\end{equation}
This quantum-controlled model for the presence or absence of a barrier also follows directly from linearity given the distinct energy distributions with and without the presence of the barrier in the two branches of the superposition. To proceed with simplifying this Hamiltonian, we use the following expansion of the field in terms of mode functions: $\hat{\phi}(x, t)~=~\sum_{m=1}^{\infty}f_m(x) \hat{b}_m e^{-i\Omega_mt} + \mathrm{h.c}$. In the transmissive branch of the superposition, the mode functions $f_m(x)$ are the standard global cavity mode functions in Eq.~\eqref{modefunctions} of the main text, such that quantisation in the transmissive branch through substitution of this expansion in Eq.~\eqref{Hphi} leads to the standard Hamiltonian of the global cavity modes $\hat{H}_T=\sum_n \Omega_n \hat{b}_n^{\dagger} \hat{b}_n$, where $\Omega_n = \frac{\pi n }{L}$. In the reflective branch of the superposition, we consider only a single mode $f_0(x)$ to fall within the reflectivity bandwidth and is made discontinuous through the barrier, we thus take as an ansatz
\begin{equation}\label{f1x2}
\begin{split}
    f_0(x)= \begin{cases}A \sin (k x), & 0<x<r \\ { B \sin (k(L-x))} & r<x<L,\end{cases}  
\end{split}
\end{equation}
The delta function potential and a single reflected frequency imply the following equations of motion (for the insant-time mode functions) in the reflective branch of the Hamiltonian in Eq.~\eqref{Hrr1} for a general time-dependent barrier strength $g$:
\begin{equation}\label{f000}
\begin{split}
    f_0''(x) - k^2f_0(x) = -2g\delta(x-r)f_0(x), \\
    \tilde{f}_n''(x) + k_n^2\tilde{f}_n(x) = 0,
\end{split}    
\end{equation}
where the second line is for $n \geq 1$ (the global cavity modes, with the lowest order mode subtracted as in Eq.~\eqref{fnx}). Integrating this from $r^+ = r - \epsilon$ to $r^- = r + \epsilon$, and taking $\epsilon \rightarrow 0$ gives the following condition on the mode function in the array's reflectivity band-width (up to surface terms):
\begin{equation}\label{f1}
    f_0'(r^+) - f_0'(r^-) = 2g f_0(r). 
\end{equation}
If $g \rightarrow 0$, then this derivative jump and continuity of $f_1$ implies algebraically that Eq.~\eqref{f1x2} reduces to the form in Eq.~\eqref{fnx}, while for sufficiently large $g$, they decouple into two completely independent sub-cavities, with boundary condition $f_1(0) = f_1(R) = f_1(r) = 0$. In order to see this, we note that continuity and the jump condition in Eq.~\eqref{f1}, respectively imply
\begin{equation}\label{jumpcontinuity}
\begin{split}
    &A \sin (k r) = B \sin (k(L-r)) \\
    &\sin (k L)+\frac{2 g}{k} \sin (k r) \sin (k(L-r))=0.
\end{split}    
\end{equation}
This implies that for perfect transmission, such that $g = 0$, we have $\mathrm{sin}(kL) = 0$, implying the mode functions patch together, and can form standing waves of the global cavity. For $g \gg k$, such that there is a strong barrier, the product term in the second equation needs to be correspondingly small which can be satisfied by either of the sine-functions vanishing sufficiently fast at the boundary. We thus expand about $kr = \pi + \epsilon_L$, and $k(L-r) = m\pi + \epsilon_R$, which for small $\epsilon_R, \epsilon_L$, gives $\sin (k r)~\propto~-\frac{k}{2g}, \quad \sin (k(L-r)) ~\propto -\frac{k}{2g}$, which for $g \rightarrow \infty$ that gives standing modes with an additional Dirichlet boundary condition at $x = r$ for the left and right sub-cavities, decoupling the cavities into independent sub-systems. More generally the second line in Eq.~\eqref{jumpcontinuity} implies two sets of solutions that require either $\sin (k r) \rightarrow 0$, $\sin (k(L-r)) \rightarrow 0$ as $g \rightarrow \infty$, by the continuity equation, one implies the other. In this limit, the normalisation constants of the global cavity mode in Eq.~\eqref{f1x2} are precisely the Bogoliubov coefficients of the N-th global mode with respect to the fundamental left and right sub-cavity modes at the same frequency - showing that the global mode reduces to the membrane-in-the-middle normal modes in the near-unit-reflectivity limit.

We next expand the field operators in terms of the mode functions. Here $\hat{a}_1$ and $\hat{\bar{a}}_k$ are left and right sub-cavity modes with the array's reflectivity band-width. Next, the total field and its conjugate momentum can be expressed as (in the Schrodinger picture)
\begin{equation}
        \hat{\phi} = \sum^\infty_{n = 0} f_n(x)\hat{q}_n,  \;  \hat{\Pi} = \sum^\infty_{n = 0} f_n(x)\hat{p}_n,
\end{equation}
where $\hat{q}_n = \frac{1}{\sqrt{2\Omega_n }}(\hat{\tilde{b}}_n^\dag + \hat{\tilde{b}}_n)$, $\hat{p}_n = -i\sqrt{\frac{\Omega_n}{2 }}(\hat{\tilde{b}}_n - \hat{\tilde{b}}_n^\dag)$, for $n \geq 1$, with 
\begin{equation}\label{fnx}
    \tilde{f}_n(x) = f_n(x) - \left(f_n \mid u_1\right)u_1(x) - \left(f_n \mid u_1^*\right)u^*_1(x) - \left(f_n \mid \bar{u}_k\right) \bar{u}_k + \left(f_n \mid \bar{u}_k^*\right) \bar{u}_k^{\dagger}.
\end{equation}
Here $f_n(x) \rightarrow \tilde{f}_n(x)$ (for $n \geq 1$) define the mode functions for the modified global cavity modes $\hat{\tilde{b}}_n$, and $f_n(x)$ are the unperturbed global cavity modes for $n \geq 1$, given in Eq.~\eqref{modefunctions} of the main text, while $u_1(x) = \frac{\theta(r-x)}{\sqrt{r \omega_1}} \sin \left(\frac{\pi x}{r}\right)$ is the fundamental mode of the left sub-cavity of length $r$, and $\bar{u}_k$ being the k-th mode of the right sub-cavity, and the inner product is with respect to the Klein-Gordon inner product. As shown in Appendix~\ref{reflappdev}, the modes $\tilde{f}_n(x)$ are orthogonal in the limit $R \gg r$, to $O(a^2)$, where $a = \frac{r}{L}$. In the limit of large $g$, the energy distributions now decouple and the left and right regions of the barrier $r$ can be independently quantised, such that the energy distribution becomes (after normal ordering)
\begin{equation}\label{HRR2}
   \hat{H}_R = \omega_1 \hat{a}_1^{\dagger} \hat{a}_1+\sum_n \Omega_n \tilde{b}_n^{\dagger} \tilde{b}_n + \bar{\omega}_k \hat{\bar{a}}_k^{\dagger} \hat{\bar{a}}_k + g \hat{\phi}_{\Delta}(r) \hat{\phi}_{\Delta}(r).
\end{equation}
This follows from breaking up the fundamental mode into two left and right sub-cavity modes, $u_1, \bar{u}_k$ and proceeding with quantisation. Importantly this re-derives the standard membrane in the middle Hamiltonian \cite{JayichAM2008Doam} from a full quantum field theory picture (further seen by Taylor expanding $g(t) =\frac{i \omega r(t)}{1 + r(t)} $ about reflectivities close to $-1$, in which the interaction term picks up squeezing and beamsplitter terms with co-efficients proportional to $\omega_1(1+r(t))$. This branch-wise quantisation procedure then produces the following final Hamiltonian deriving the Hamiltonian given in the main text:
\begin{equation}
    \hat{H} = \hat{H}_T \otimes|T\rangle_c\left\langle\left. T\right|_c\right. + \hat{H}_R \otimes|R\rangle_c\left\langle\left. R\right|_c\right..
\end{equation}
If we now expand the frequency-restricted field operator $\hat{\phi}_{\Delta}(r)$, in terms of global-cavity  modes, and using for large $g$, that the mode functions scale as $\frac{k}{g}$. Now, taking $g(t)$ as above, where $r_\mathrm{eff}(t)$ is the effective reflectivity and expanding for reflectivities close to zero. In the small $g$ limit, the energy distribution becomes:
\begin{equation}\label{HRR2}
   \hat{H}_R = \Omega_N\hat{b}^\dag_N\hat{b}_N +  \sum_n  \Omega_{n \neq N} \hat{b}_n^{\dagger} \hat{b}_n +\Omega_N r_\mathrm{eff}(t) \hat{\phi}_{\Delta}(r) \hat{\phi}_{\Delta}(r),
\end{equation}
where $\Omega_N = \frac{\pi N}{L}$ is the global cavity mode $N$, and now expanding $\hat{\phi}_\Delta$ in terms of global cavity mode $f_N$. In particular the continuity condition in Eq.~\eqref{f1x2} implies that the distinct energy distributions patch together, and can be diagonalised as $\Omega_N\hat{b}^\dag_N\hat{b}_N$.

\subsection{Non-adiabatic corrections}\label{multinon}
Here as a further consistency check, we examine how for slow but finite switching rates, we obtain non-adiabatic corrections to the adiabatic theorem which still yields measurable photon content. We further complete the derivation of the quantum-controlled Hamiltonian using full QED.

\subsubsection{Scalar field model}
Starting from the ground state of the global cavity, this typically implies that local sub-cavities will not show any particle content. In this Appendix, we show that this switching time still leads to non-adiabatic corrections that generate particle content in the local sub-cavities. Specifically, we examine non-adiabatic corrections to the adiabatic theorem: that states that for slowly varying Hamiltonians the eigenstates track the Hamiltonian at each instant of time. Now considering a time-dependent reflection co-efficient, the interaction term in Eq.~\eqref{HRR2} becomes time-dependent:
\begin{equation}
H_R(t) =  \omega_1 \hat{a}_1^{\dagger} \hat{a}_1 + \bar{\omega}_k \hat{\bar{a}}_k^{\dagger} \hat{\bar{a}}_k + \sum_n\Omega_n \tilde{b}_n^\dagger \tilde{b}_n + \hat{H}_\mathrm{int}(t),
\end{equation}
Here, non-adiabatic corrections to the adiabatic theorem yield the amplitude to find the field excited to a higher energy state (to transition to instantaneous Hamiltonian eigenstate):
\begin{equation}
    c^{(1)}_{1,k}(T) = - \int_0^T \mathrm{d}t \frac{\bra{m(t)}  \dot{H}(t) \ket{n(t)}}{E_n(t) - E_m(t)}e^{i\phi_{nm}(t)}, 
\end{equation}
where $T$ is the time-scale of the slow-switching, and $c^{(1)}_{1,k}$ is the amplitude for pairs of particles to be created across the sub-cavities by the two-mode squeezing interactions, in essence to transition to the $n = 1$ Fock state, due to squeezing. Here, $\phi_{nm}(t) = \theta_n(t) - \theta_m(t)$ and $\theta_n(t) \equiv -\frac{1}{\hbar} \int_0^t E_n\left(t^{\prime}\right) d t^{\prime}$. For $\Gamma T \ll 1$, with the exponential rise of the reflectivity according to the linewidth, this produces $c_{1,k}^{(1)}(T) \sim r_\mathrm{max}(\frac{\Gamma_e}{\omega_1})$ being the leading corrections for the sub-cavity mode to transition due to non-adiabatic contributions. The power law suppression with the slow-switching switching time corroborates the predictions of the toy model for the scaling of the suppression of particle creation effects (in the sense of a power-law suppression), and reduces to the previously studied single-mode-case in the relevant regime.

 We firstly extend the scalar field model (associated one-to-one with the Coulomb gauge Hamiltonian) to the case of multiple modes to explicitly solve for the particle creation. If we take the homogenous solution for the previously derived scalar field equation for a delta function barrier, now matching the reflectivity co-efficient derived from full microscopic QED $r(t)=\left(1-e^{-\Gamma_e t / 2}\right)$ to the permittivity, 
\begin{equation}\label{fxt222}
\partial_x^2 f(x, t)+k_0^2 f(x, t)=\frac{r(t) k_0}{r(t) + 1} f(x, t) \delta\left(x-r_x\right),
\end{equation}
now taking, $\widehat{E}(x, t)=\sum_n \hat{q}_n(t) u_n(x, t)$, with $\left.\partial_t^2 \widehat{E}=\sum_n \ddot{\hat{q}}{ }_n u_n + 2\dot{\hat{q}}_n \dot{u}_n+\hat{q}_n \ddot{u}_n\right)$, as for the scalar field model - mapped one to one from scalar field theory to Quantum Electrodynamics, and expanding in the full time-dependent wave equation for the sourced electric field. Now with,
\begin{equation}
\begin{split}
    M_{m n}(t) &\equiv \int d x u_m(x, t) \dot{u}_n(x, t) \\
  K_{m n}(t)  &= \int d x u_m(x, t)\ddot{u}_n(x, t) ,
\end{split}
\end{equation}
we obtain the equation involving a time-dependent basis structure,
\begin{equation}
\ddot{\hat{q}}_m+\omega_m^2(t) \hat{q}_m=- \sum_n M_{m n}(t)\dot{\hat{q}}_n-\sum_n K_{m n}(t) \hat{q}_n.
\end{equation}
Now, the mode-mixing terms for $n \neq m$ are suppressed by the free-spectral-range compared to the effect of the $n = m$ terms. The modes here correspond to the modes of the instant-in-time modes in Eq.~\eqref{fxt222}. 

\subsubsection{Quantum Electrodynamics}\label{EPP}
We take as a starting point the Hamiltonian that we solved for in Eq.~\eqref{timed1} after eliminating the atomic degrees of freedom. If we define the instantaneous creation and annihilation operators (this follows with-out using the delta barrier model matched from a scalar field theory - if we use the full QED Hamiltonian in Eq.~\eqref{timed1} and define the following re-scaled time-dependent annihilation operators in terms of the instantaneous mode functions that diagonalise the time-dependent Hamiltonian in Eq.~\eqref{timed1} in terms of the static fixed cavity modes), we can now diagonalise this non-hermitian Hamiltonian in terms of a time-dependent basis structure through a bi-orthogonal transformation \cite{PhysRevA.96.062130} , 
\begin{equation}
\begin{aligned}
    \binom{b}{\bar{b}^{\dagger}}=\mathcal{W}\binom{a}{a^{\dagger}}, \quad \mathcal{W}=\left(\begin{array}{cc}
u & v \\
\bar{v}^* & \bar{u}^*
\end{array}\right)    
\end{aligned}
\end{equation}
taking the time derivative where $[b, \bar{b}]=u \bar{u}^* - v \bar{v}^*=1$, 
\begin{equation}
\begin{split}
\binom{\dot{b}}{\bar{b}^{\dagger}} &= \dot{\mathcal{W}}\mathcal{W}^{-1}\binom{b}{\bar{b}^{\dagger}}+\mathcal{W}\binom{\dot{a}}{\dot{a}^{\dagger}}  \\
\binom{\dot{b}}{\bar{b}^{\dagger}} &= \dot{\mathcal{W}}\mathcal{W}^{-1}\binom{b}{\bar{b}^{\dagger}}+\mathcal{W}i\left[H,\binom{a}{a^{\dagger}}\right] \\
\binom{\dot{b}}{\bar{b}^{\dagger}} &= \dot{\mathcal{W}}\mathcal{W}^{-1}\binom{b}{\bar{b}^{\dagger}}+\mathcal{W}\mathcal{W}^{-1}i\left[H,\binom{b}{\bar{b}^{\dagger}}\right]\\
\binom{\dot{b}}{\bar{b}^{\dagger}} &= \dot{\mathcal{W}}\mathcal{W}^{-1}\binom{b}{\bar{b}^{\dagger}}+ i\Omega_n(t)\binom{-b}{\bar{b}^{\dagger}},
\end{split}
\end{equation}
where going to the second line, we have used that for this specific non-hermitian hamiltonian that it generates time-evolution of the individual creation and annihilation operators through the commutator - the other relations follow from Ref.~\cite{PhysRevA.96.062130}. We thus obtain the equation of motion of a harmonic oscillator with a time-dependent frequency (to lowest order in $\frac{\dot{\Omega}_n(t)}{ \Omega_n(t)}$, and again having using relations from Ref.~\cite{PhysRevA.96.062130} when converting from the basis of time-independent static modes $\hat{a}$),
\begin{equation}\label{annn}
\begin{split}
        \dot{b} &\approx -i \Omega_n(t) \hat{b}(t) + \frac{\dot{\Omega}_n(t)}{4 \Omega_n(t)} \bar{b}^\dagger(t)  -i\hat{\eta}_\mathrm{in}(t), 
\end{split}
\end{equation}
where we now have an effective input noise term with effective decay rate (enhanced due to the collective coupling and dropping the counter-rotating noise piece),
\begin{equation}
\begin{split}
        \kappa_\mathrm{eff} &= g_q^2 \frac{N\Gamma_e}{\Gamma_e^2 + \Delta_e^2}.   
\end{split}   
\end{equation}
on the photonic mode due to the coupling to the atomic polarisation operator which manifests in the correlator  $\left\langle\hat{\eta}_{\mathrm{in}}(t) \hat{\eta}_{\mathrm{in}}^{\dagger}\left(t^{\prime}\right)\right\rangle \approx  \kappa_{\mathrm{eff}}\delta\left(t-t^{\prime}\right)$. Solving the particle production purely from the connection piece we have the analytical solution $\langle \hat{n} \rangle \sim \left(\frac{\Gamma_e}{n\Omega_n}\right)^2$, consistent with the scaling computed from the scalar field model, with an additional factor corresponding to the global cavity mode number $n$. This is computed firstly by solving the Heisenberg Langevin equation with solution for the squeezing:
\begin{equation}\label{att1}
    b(t) \approx b(t_0)e^{-\int_0^t i\omega(\tau)\mathrm{d\tau}} +  \beta(t)\bar{b}^\dagger(t_0) + \mathrm{noise} \; \mathrm{terms}, 
\end{equation}
where $\beta(t) \approx \int_0^T\mathrm{dt}\frac{\dot{\Omega}_n(t)}{4 \Omega_n(t))}e^{i\int^t_0 \Omega(\tau) \mathrm{d\tau}}$ and we solve for the time-dependent frequency by taking the total derivative of the quantisation condition in Eq.~\eqref{jumpcontinuity} and expanding about reflectivities close to unity. We can further use the Hamiltonian in Eq.~\eqref{timed1}, which with time-dependent mode functions now implies the exact same solution.

Integrating Eq.~\eqref{annn} and restoring the implicit projector onto the control atom state, then combining with the cavity energy distribution in the transmissive branch, recovers the effective main-text Hamiltonian in the weak-drive limit. Equivalently, this manifests as a frequency shift of the control: the Heisenberg evolution of the weakly driven coherence $\sigma_{TR}$ acquires a term proportional to the frequency difference between the two branches:
\begin{equation}
    \dot{\hat{\sigma}}^c_{TR}  = -i(\nu + \hat{H}_R -  \hat{H}_T)\hat{\sigma}^c_{TR} -ig\hat{\sigma}^c_z,
\end{equation}%
where we have that $\hat{H}_R$ is the effective Hamiltonian of the photonic cavity, conditioned on the reflective branch of the superposition, after the previously discussed adiabatic elimination procedure in Eq.~\eqref{timed1}, and $\hat{H}_T$ is the Hamiltonian of the global cavity in the transmissive branch of the superposition for which the photonic modes are unperturbed, and $\hat{\sigma}^c_z$ is the Pauli-z matrix on the control atom with transition frequency $\nu$. Using the full solution in Eq.~\eqref{att1}, for the annihilation operator in the reflective branch, such that $\hat{H}_R = \Omega(t)\bar{b}^\dagger(t)^\dag b(t)$ (with $\Omega(t)^2 \propto \omega_0^2 + \frac{\omega_0^2\chi(r, t)^2}{2}$, where $\chi(r, t)~=~f(r, t)\left(1-e^{-\Gamma_e t}\right)$) this derives from full first-principles quantum electrodynamics the frequency shift on the control from the microscopic interactions between the array atoms and the field modes. In order to see this explicitly, we define the operator valued frequency:
\begin{equation}
  \hat{\Omega}(t) = \nu + \hat{H}_R(t) -  \hat{H}_T,
\end{equation}
with $\ket{T}$ state initialised control atom, we obtain $\widehat{\sigma}_{T R}^c(t) \approx-g \frac{e^{-i \widehat{\Omega} t}-1}{\widehat{\Omega}}$ as the solution for the coherence. Once this is propagated into the solution for the probability for the control to trnasition (in the regime for which the array response function is much faster than the coupling strength to the control atom, we can use an adiabatic approximation and pull-out the stead state frequency operator after the switch), we obtain:
\begin{equation}
    P_R(t) = 4g^2\mathrm{Tr}_\mathrm{F}\left[\frac{\mathrm{sin}^2(\hat{\Omega}(t_f)t)}{\hat{\Omega}(t_f)^2}\right],
\end{equation}
where $t_f$ is some fixed time $t_f \gg \frac{1}{\Gamma_e}$ and the trace refers to the trace over the photonic field states. For an initial field state being the vacuum this pulls out the required frequency shift.

\subsection{Alternative Reflective Hamiltonian Derivation}\label{reflappdev}
\begin{figure}
    \centering
    \includegraphics[width=0.9\linewidth]{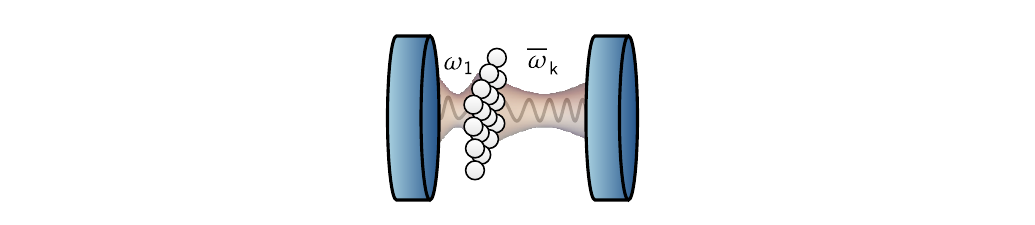}
    \caption{An atom array in a photonic cavity that is highly reflective within some frequency band-width $\Delta$, can trap standing waves of frequency $\omega_1 \in \Delta$ of the left sub-cavity, and $\bar{\omega}_k \in \Delta$ of the right sub-cavity. The Hamiltonian for an atom array that is highly reflective in this frequency band-width can be constructed as the sum of free Hamiltonian terms of the left and right sub-cavities, including vacuum terms.  }
    \label{fig:atom2}
\end{figure}

In this section, we construct an alternative derivation for the Hamiltonian for a cavity containing a reflective membrane that reflects only a single frequency by considering the allowed mode of the cavity field, which is relevant for the discussion at the end of Appendix~\ref{beam1}. This version only applies in the fast-switching regimne. In our quantum-controlled model, the Hamiltonian we construct here will correspond to the Hamiltonian of the photonic cavity when the atomic array is in its reflective state. We first consider the decomposition of the modes of a photonic cavity into its sub-regions, as described in Ref.~\cite{BrownEricG.2015Wdim}. The global cavity modes ($\hat{b}_n$) are related to the modes of the left and right sub-cavities ($\hat{a}_l$ for the left sub-cavity and $\hat{\bar{a}}_l$ for the right sub-cavity) through the following Bogoliubov transformation, which gives the same Bogoliubov co-efficients as the inverse transformation in Eq.~\eqref{bogoaaa}:
\begin{equation}\label{bogoatob}
	\hat{b}_n = \sum_l \alpha_{ln}\hat{a}_l - \beta_{ln}\hat{a}_l^{\dagger} + \bar{\alpha}_{ln}\hat{\bar{a}}_l -\bar{\beta}_{ln}\hat{\bar{a}}_l^{\dagger},  
\end{equation}
To isolate the contribution of a specific set of modes that are reflected by the array, we subtract the terms corresponding to those modes from the global mode expression. Specifically, by subtracting the contributions of the $\hat{a}_1$ mode (from the left sub-cavity) and the set of modes $\hat{\bar{a}}_k$, with frequencies within the reflectivity bandwidth, $\bar{\omega}_k\in \Delta$ (from the right sub-cavity), we define a modified global mode $\tilde{b}_n$ as:  
\begin{equation}\label{redefinedmodes}
\begin{split}
\tilde{b}_n &= \hat{b}_n -  \alpha_{1n}\hat{a}_1 + \beta_{1n}\hat{a}_1^{\dagger}  + \sum_k - \bar{\alpha}_{kn}\hat{\bar{a}}_k + \bar{\beta}_{kn}\hat{\bar{a}}_k^{\dagger} \\
&=\sum_{l \geq 2} \alpha_{l n} \hat{a}_l-\beta_{l n} \hat{a}_l^{\dagger}+\sum_{l \neq k} \bar{\alpha}_{l n} \hat{\bar{a}}_l-\bar{\beta}_{l n} \hat{\bar{a}}_l^{\dagger}.
\end{split}
\end{equation}
This  explicitly subtracts the contributions of the $\hat{a}_1$ and $\hat{\bar{a}}_k$ modes, thereby modifying the global mode $\hat{b}_n$ to exclude their contributions. If we now expand the global Hamiltonian in terms of each of the sub-cavity modes, and isolate out the contributions from the single frequency sub-cavity modes $\hat{a}_1$ and $\hat{\bar{a}}_k$ (this represents that when the mirror is off-centre, in general higher order modes $k \geq 1$ can be the frequency which is filtered by the mirror) we obtain
\begin{equation}\label{Hinitexp}
\begin{split}
 \hat{H}_T &\rightarrow \sum_{n} \bigg[ \Omega_n\tilde{b}_n^{\dagger}\tilde{b}_n + \Omega_n\alpha_{1n}(\tilde{b}_n\hat{a}_1^{\dagger} + \hat{a}_1\tilde{b}_n^{\dagger}) - \Omega_n\beta_{1n}(\hat{a}_1\tilde{b}_n + \tilde{b}_n^\dag\hat{a}_1^{\dagger}) + \Omega_n\left((\alpha_{1n}^2 + \beta_{1n}^2 )\hat{a}_1^{\dagger}\hat{a}_1\right)  \\ 
    & \qquad   - \Omega_n \alpha_{1n}\beta_{1n}(\hat{a}_1^{\dagger}\hat{a}_1^{\dagger} + \hat{a}_1\hat{a}_1) + \Omega_n\beta_{1n}^2 \bigg] - \sum_k  \bigg[ \Omega_n \bar{\beta}_{kn}\left(\hat{\bar{a}}_k \tilde{b}_n+\tilde{b}_n^{\dagger} \hat{\bar{a}}_k^{\dagger}\right) - \Omega_n\bar{\alpha}_{kn}(\tilde{b}_n^{\dagger}\hat{\bar{a}}_k + \hat{\bar{a}}_k^{\dagger}\tilde{b}_n)
    \vphantom{\sum_n} 
     \\
    & \qquad + \Omega_n(\alpha_{1 n}\bar{\alpha}_{k n} + \bar{\beta}_{k n}\beta_{1 n})(\hat{a}_1^{\dagger}\hat{\bar{a}}_k + \hat{a}_1\hat{\bar{a}}_k^{\dagger})  - \Omega_n(\alpha_{1 n}\bar{\beta}_{k n} + \bar{\alpha}_{k n}\beta_{1 n} ) (\hat{a}_1^{\dagger} \hat{\bar{a}}_k^{\dagger} + \hat{a}_1 \hat{\bar{a}}_k) 
    \vphantom{\bigg] \sum_n} 
    \\
    & \qquad  +\Omega_n\left(\left(\bar{\alpha}_{k n}^2+\bar{\beta}_{k n}^2\right)\hat{\bar{a}}_k^{\dagger} \hat{\bar{a}}_k\right)-\Omega_n \bar{\alpha}_{k n} \bar{\beta}_{k n}\left(\hat{\bar{a}}_k^{\dagger} \hat{\bar{a}}_k^{\dagger}+\hat{\bar{a}}_k \hat{\bar{a}}_k\right)+\Omega_n \bar{\beta}_{k n}^2 \bigg] .
    \vphantom{\sum_n} 
\end{split}
\end{equation}
The presence of the single frequency mirror removes the interactions between the $\hat{a}_1$ mode and all right sub-cavity modes, as well as all $\hat{\bar{a}}_k$ modes and all left sub-cavity modes. This follows how for the perfectly reflecting mirror considered in Sec.\ \ref{subregpartcont}, the effect of the mirror is to remove interactions and squeezing terms across the sub-cavity modes that are present in the global cavity Hamiltonian. Applying the same principle, but restricted to the set of frequencies reflected by the array (which was derived rigorously in Appendix~\ref{beam1}), we obtain,
\begin{equation}\label{Hbn4}
\begin{split}
    \hat{H}_R &= \omega_1 \hat{a}_1^{\dagger} \hat{a}_1 + \sum_{n} \bigg[\Omega_n \tilde{b}_n^{\dagger} \tilde{b}_n + \Omega_n\beta_{1n}^2 \bigg] + \sum_{k} \bigg[ \bar{\omega}_k \hat{\bar{a}}_k^{\dagger} \hat{\bar{a}}_k \bigg] +  \sum_{k,n} \bigg[\Omega_n\bar{\beta}_{k n}^2\bigg] , \\
    &\simeq \omega_1 \hat{a}_1^{\dagger} \hat{a}_1 + \sum_{n} \Omega_n \tilde{b}_n^{\dagger} \tilde{b}_n + \sum_{k}\bar{\omega}_k \hat{\bar{a}}_k^{\dagger} \hat{\bar{a}}_k ,
\end{split}
\end{equation}
where the sum over $k$ extends to all right sub-cavity modes that the single frequency mirror reflects (all modes that fall within the linewidth of the atomic array's reflectivity), and in the second line we have taken the limit $L \gg r$, where $r$ is the size of the sub-cavity, and neglected all other terms as they are sub-leading in this regime. Importantly, in this regime, the dominant frequency shift between the transmissive and reflective states comes from difference in particle content, rather than constant offsets $\Omega_n \beta_{1 n}^2, \Omega_n \bar{\beta}_{k n}^2$. The Hamiltonian of Eq.~\eqref{Hbn4} represents standing waves at the frequency $\omega_1$ in the left sub-cavity, with standing waves of frequencies 
$\bar{\omega}_k$ 
in the right sub-cavity as displayed in Fig.~\ref{fig:atom2}. 
We now construct the Hamiltonian of the global cavity with the sub-cavities removed as
\begin{equation}
\begin{split}
 \hat{H}_{\perp} &= \sum_n \Omega_n\tilde{b}_n^{\dagger}\tilde{b}_n,
    \vphantom{\sum_n}
\end{split}
\end{equation}
which is completely orthogonal to the individual single frequency sub-cavities as 
\begin{equation}
    [\hat{H}_{\perp}, \omega_1 \hat{a}_1^{\dagger} \hat{a}_1 ] = [\hat{H}_{\perp}, \bar{\omega}_1 \hat{\bar{a}}_1^{\dagger} \hat{\bar{a}}_1 ] = 0.
\end{equation}
We therefore, use the following Hamiltonian for the single frequency reflective mirror in the $L \gg r$ regime,
\begin{equation}\label{reflectH}
     \hat{H}_R =  \omega_1\hat{a}_1^{\dagger} \hat{a}_1 + \sum_{n} \Omega_n\tilde{b}_n^{\dagger}\tilde{b}_n  + \sum_{k} \bar{\omega}_k\hat{\bar{a}}_k^{\dagger} \hat{\bar{a}}_k,
\end{equation}
noting that the commutator 
\begin{equation}
\begin{split}
    [\tilde{b}_i, \tilde{b}_j^\dag] &= \delta_{ij} + \alpha_{1i}\alpha_{1j}^* - \beta_{1i}\beta_{1j}^* + \sum_k \bar{\alpha}_{ki}\bar{\alpha}_{kj}^* - \bar{\beta}_{ki}\bar{\beta}_{kj}^*, \\
    = \delta_{ij} &+ \frac{\sqrt{ij}(-1)^{i+j}\sin(i\pi a)\sin(j\pi a)}{\pi^2(aj-1)(ai-1)} - \frac{\sqrt{ij}(-1)^{i+j}\sin(i\pi a)\sin(j\pi a)}{\pi^2(aj+1)(ai+1)} \\
    &+ \frac{k\sin(i\pi a)\sin(j\pi a)}{\sqrt{ij}\pi^2(\bar{a}'j-k)(\bar{a}'i-k)} - \frac{k\sin(i\pi a)\sin(j\pi a)}{\sqrt{ij}\pi^2(\bar{a}'j+k)(\bar{a}'i+k)}
\end{split}
\end{equation}
in the regime $L \gg r$, the commutator approaches $[\tilde{b}_i, \tilde{b}_j^\dag] = \delta_{ij} + O(a^2)$, and we obtain a division of the sub-cavities in terms of purely orthogonal modes in this limit. With this form of the reflective Hamiltonian, the full quantum-controlled Hamiltonian in the main text follows from linearity. Alternatively, we can also use Eq.~\eqref{reflectH}, as an order of magnitude estimate for the $L \sim r$ regime, as the interaction terms between the modified global cavity modes $\tilde{b}_n$ and the sub-cavity modes are sub-leading even in this regime compared to the free-energy terms, with the largest interaction terms being $O(0.5\Omega_n)$. Specifically we convert to the rotating frame of the free-energy terms and drop these terms as counter-rotating terms. Therefore, we use this Hamiltonian as the core model for all positions of the array through-out the cavity, however, the accuracy approaches unity for the $L \gg r$ regime.

\section{Perturbative Dynamics} 

\subsection{Analytical Estimate for the ideal frequency shift}\label{analytical111}
In this section we derive the analytical estimate for the frequency shift in the control due to particle-creation from the vacuum. The full Hamiltonian in the rotating frame of the control atom is
\begin{equation}
    \hat{H}=\hat{H}_{\mathrm {cav }}+\hat{H}_{\mathrm {switch }},
\end{equation}
where,
\begin{equation}
    \begin{split}
          \hat{H}_{\mathrm {cav }} &= \hat{H}_T \otimes|T\rangle\langle T|+ \hat{H}_{\mathrm{R}}  \otimes|R\rangle\langle R| 
          \vt \\   
          \hat{H}_{\mathrm {switch }} &= g\left(|T\rangle\langle R| e^{i\left(\omega_D-\nu\right) t}+\mathrm{ h.c }\right), 
          \vt 
    \end{split}
\end{equation}
and we recall that $\hat{H}_T = \sum_n \Omega_n \hat{b}_n^{\dagger} \hat{b}_n$ and $\hat{H}_R=\omega_1 \hat{a}_1^{\dagger} \hat{a}_1+\sum_n \Omega_n \tilde{b}_n^{\dagger} \tilde{b}_n+\sum_k \bar{\omega}_k \hat{\bar{a}}_k^{\dagger} \hat{\bar{a}}_k$. Now, we can express $\hat{a}_1^\dag\hat{a}_1$ in terms of the global modes as
\begin{equation}
   \omega_1\hat{a}_1^\dag\hat{a}_1 =  \sum_n\left(\omega_n \hat{b}_n^{\dagger} \hat{b}_n-g_n \hat{b}_n^{\dagger} \hat{b}_n^{\dagger}-g_n^* \hat{b}_n \hat{b}_n + \omega_1\left|\beta_{1, n}\right|^2 \right) + \sum_{n \neq m}\left(f_{n, m} \hat{b}_n^{\dagger} \hat{b}_m-g_{n, m} \hat{b}_n^{\dagger} \hat{b}_m^{\dagger}-g_{m, n}^* \hat{b}_n \hat{b}_m+f_{n, m}^* \hat{b}_n \hat{b}_m^{\dagger}\right).
\end{equation}
where, 
\begin{equation}
\begin{aligned}
    \omega_n &= \omega_1\left(\left|\alpha_{1, n}\right|^2+\left|\beta_{1, n}\right|^2\right), 
\vt \\
    f_{n, m} &= \omega_1 \alpha_{1, n} \alpha_{1, m}^\star, 
\vt 
    \\
g_n &= \omega_1 \alpha_{1, n} \beta_{1, n}^\star, 
\vt 
\\
g_{n, m} &= \omega_1 \alpha_{1, n} \beta_{1, m}^\star.
\vt 
\end{aligned}
\end{equation}
We now convert into the rotating frame of $\hat{H}_{\mathrm{T}} \otimes|R\rangle\langle R|$, and drop all of the counter-rotating terms, and then when converting back out of the rotating frame, we are left with the $\sum_n\omega_n \hat{b}_n^{\dagger} \hat{b}_n$ term. This produces the Hamiltonian:
\begin{equation}
    \hat{H}_{\mathrm {cav }} = \hat{H}_T \otimes|T\rangle\langle T|+\left(\sum_n\left(\omega_n\hat{b}_n^{\dagger} \hat{b}_n + \Omega_n\hat{b}_n^{\dagger} \hat{b}_n + \Omega_n \tilde{b}_n^{\dagger} \tilde{b}_n  + \omega_1\left|\beta_{1, n}\right|^2\right) +\sum_k \bar{\omega}_k \hat{\bar{a}}_k^{\dagger} \hat{\bar{a}}_k  \right) \otimes|R\rangle\langle R|,
\end{equation}
and, as derived previously, after conversion to the interaction picture, 
\begin{equation}
P_R=g^2 \int_0^t \mrm{d} t^{\prime} \mrm{d} t^{\prime \prime}\left\langle 0_T\right| e^{i \hat{H}_R\left(t^{\prime}-t^{\prime \prime}\right)}\left|0_T\right\rangle e^{i\left(\omega_D-\nu\right)\left(t^{\prime}-t^{\prime \prime}\right)}
\end{equation}
gives the probability for the control atom to transition from the transmissive to the reflective state. The analytical estimate for the frequency shift 
\begin{equation}
    \delta_R = \sum_n \omega_1\left|\beta_{1, n}\right|^2,
\end{equation}
 provides a good approximation of the precise numerical solution. We have omitted the contribution from the $\sum_n \tilde{b}_n^{\dagger} \tilde{b}_n$ term in the Hamiltonian, as we numerically evaluate its contribution to be sub-leading, while the dominant contribution to the frequency shift evaluates to be $\sum_n \left|\beta_{1, n}\right|^2 \simeq 0.075$, in the regime $L \gg r$. This is the frequency shift which we use in the main text, and serves as an order of magnitude approximation in the regime for which $L \sim r$, as the additional interaction terms are sub-leading in this regime, which we confirmed through a numerical diagonalisation including all terms.

\subsection{Perturbative Dynamics of a Quantum-Controlled Photonic Cavity}\label{perturblim}
Here we present a perturbative analytical solution of the dynamics of a quantum-controlled photonic cavity. To derive this, we consider the full Hamiltonian, including the free Hamiltonian of the control:
\begin{equation}
\begin{split}
    \hat{H} &= \frac{\hbar\nu}{2}(\ket{R}\bra{R} -  \ket{T}\bra{T}) +  \hat{H}_T \otimes \ket{T}\bra{T} + \hat{H}_R \otimes \ket{R}\bra{R} + g(\ket{T}\bra{R}e^{+i\omega_D t} + \ket{R}\bra{T}e^{-i\omega_D t}),
    \vt 
\end{split}  
\end{equation}
which includes a drive applied on the control atom of strength $g$ and frequency $\omega_D$. Now transforming to the rotating frame of the control, produces:
    \begin{align}
        \h H &\to e^{i \h H_\mrm{free} t/\hslash} \h H e^{-i \h H_\mrm{free} t/\hslash} 
        \vt = \h H_T \otimes | T \rangle\langle T | + \h H_R \otimes | R \rangle\langle R | + g ( | T \rangle\langle R | e^{i(\omega_D - \nu)t} + | R \rangle\langle T | e^{i ( \nu - \omega_D ) t} ), 
    \end{align}
    where $\h H_\mrm{free} = (\hslash\nu/2) ( | R \rangle\langle R | - | T \rangle\langle T | )$ is the free Hamiltonian of the control. We now break up the following contributions to the Hamiltonian:
\begin{equation}
    \begin{split}
        \hat{H}_0 &=  \hat{H}_T \otimes \ket{T}\bra{T} + \hat{H}_R \otimes \ket{R}\bra{R}
        \vt \\
        \hat{H}_g &= g(\ket{T}\bra{R}e^{i(\omega_D - \nu)t} + \ket{R}\bra{T}e^{-i(\omega_D - \nu)t}).
        \vt 
    \end{split} 
\end{equation}
We treat $\hat{H}_g$ as a perturbation, and work in the interaction picture:
\begin{equation}
\begin{split}
    \hat{H}_g(t) &= e^{i \hat{H}_0 t} \hat{H}_g e^{-i \hat{H}_0 t}.
    \vt 
\end{split}
\end{equation}
we have $|\psi(t)\rangle_I=e^{i \hat{H}_0 t}|\psi(t)\rangle_{S} = U_I(t)\left|\psi_I(0)\right\rangle$, where 
\begin{equation}
\begin{split}
    U_I(t) &= \mathcal{T}\left\{e^{-i \int_0^t d t^{\prime} \hat{H}_g\left(t^{\prime}\right)}\right\} = \hat{\mathds{I}}-i \int_0^t d t^{\prime} \hat{H}_g\left(t^{\prime}\right)-\frac{1}{2} \int_0^t d t^{\prime} \int_0^{t^{\prime}} d t^{\prime \prime} \hat{H}_g\left(t^{\prime}\right) \hat{H}_g\left(t^{\prime \prime}\right)+\ldots,
\end{split}
\end{equation}
where $\mathcal{T}$ denotes time-ordering. Including terms up to second order in $g$, gives explicitly
\begin{equation}
\begin{split}
        \hat{U}_I^{(2)} &= \hat{\mathds{I}} - i g \int_0^t \mathrm{d} t' \: \big( e^{i\hat{H}_Tt'} | T \rangle\langle R | e^{-i\hat{H}_R t'} e^{i(\omega_D-\nu)t'} + \mathrm{h.c} \big) 
         \\
        & \qquad - \frac{g^2}{2} \int_0^t\mathrm{d}t' \int_0^{t'} \mathrm{d} t'' \: \big( e^{i\hat{H}_Tt'} | T \rangle\langle R | e^{-i\hat{H}_Rt'} e^{i(\omega_D-\nu)t' } + \mathrm{h.c} \big) \big( e^{i\hat{H}_Tt''} |T  \rangle\langle R | e^{-i\hat{H}_Rt''} e^{i(\omega_D-\nu)t''} + \mathrm{h.c} \big) + O ( g^2 )  
         \\
        &= \hat{\mathds{I}} - i g \int_0^t \mathrm{d} t' \: \big( e^{i\hat{H}_T t'} | T \rangle\langle R | e^{-i\hat{H}_R t' } e^{i(\omega_D - \nu) t'} + \mathrm{h.c} \big) 
         \\
        & \qquad - \frac{g^2}{2} \int_0^t\mathrm{d} t' \int_0^{t''} \mathrm{d} t'' \big( e^{i\hat{H}_Tt'} | T \rangle\langle 
        R | e^{-i \hat{H}_Rt'} e^{i\hat{H}_Tt''} | T \rangle\langle R | e^{-i\hat{H}_Rt''} e^{i(\omega_D- \nu ) ( t' + t'' ) } 
         \\
        & \qquad \qquad + e^{i\hat{H}_T t'} |T \rangle\langle R | e^{-i\hat{H}_Rt'} e^{i \hat{H}_Rt''} | R \rangle\langle T | e^{-i\hat{H}_Tt''} e^{i(\omega_D - \nu ) (t' - t'')} 
        \vphantom{\int_0^t}
        \\
        & \qquad \qquad + e^{i\hat{H}_Rt'} | R \rangle\langle T | e^{-i \hat{H}_Tt'} e^{i \hat{H}_Tt''} | T \rangle\langle R | e^{-i\hat{H}_Rt''} e^{-i(\omega_D - \nu )(t' - t'' ) } 
        \vphantom{\int_0^t}
         \\
        & \qquad \qquad + e^{i\hat{H}_Rt'} | R \rangle\langle T | e^{-i\hat{H}_Tt'} e^{i \hat{H}_Rt''} | R \rangle\langle T | e^{i \hat{H}_Tt''} e^{-i(\omega_D - \nu )(t' + t'')} 
         ) + O ( g^3 ) .
        \vphantom{\int_0^t}
        \label{eq15}
    \end{split}
    \end{equation}
In the main text, this unitary is restricted to $O(g)$ terms relevant for the leading order computations of the transition rate of the control atom. If the control is initially in the transmissive state, the probability for it to flip to the reflective state after unitary evolution up to $O(g)$ is
\begin{equation}
P_R=g^2 \int_0^t \mrm{d} t^{\prime} \mrm{d} t^{\prime \prime}\left\langle 0_T\right| e^{i \hat{H}_R\left(t^{\prime}-t^{\prime \prime}\right)}\left|0_T\right\rangle e^{i\left(\omega_D-\nu\right)\left(t^{\prime}-t^{\prime \prime}\right)},
\end{equation}
which is used in the main text. The higher order terms correspond to Lamb-shift-type energy corrections.

\subsection{Imperfect Reflectivity}\label{app:E}
Here, we outline a toy model for switching on a mirror of imperfect reflectivity. Our model for a mirror with imperfect reflectivity is adapted from Ref.~\cite{BrownEricG.2015Sasc}, which considers particle content due to the presence of a time-dependent boundary condition at the origin of Minkowski spacetime, as well as that of a cavity witih boundaries at $x = \pm a/2$. A similar problem was considered in Ref.\ \cite{Foo_2020}, for a mirror following conformal diamond time with tunable reflectivity.  

As elsewhere in this manuscript, we consider a real massless scalar field $\hat \phi \equiv \hat \phi(t,x)$. Introducing a mirror at $x = 0$ modifies the scalar Klein-Gordon equation to read, 
\begin{align}
    \p_t^2 \hat \phi - \Delta_{\theta(t)} \hat \phi = 0 
    \vt 
    \label{eq30}
\end{align}
where $\{ - \Delta_{\theta} | \theta \in [ 0 , \pi/2] \}$ is the one-parameter family of self-adjoint extensions of $- \p_x^2$ on $L^2(\mathds{R}$ \textbackslash $\{0\} )$. The limiting cases $\theta(t) = 0, \pi/2$ correspond to a perfectly reflective and transmissive boundary respectively, with intermediate values interpolating between the two. One can write Eq.\ (\ref{eq30}) as, 
\begin{align}
    \bigg[ \p_t^2 - \p_x^2 + \frac{2 \cot ( \theta ( t ) ) }{L} \delta(x) \bigg] \hat \phi = 0 
    \vt 
    \label{eq33}
\end{align}
where $L$ has dimensions of length, which we henceforth set to unity. The presence of the mirror thus acts as a potential term proportional to $\delta(x)$ with a time-dependent coefficient, tending to zero as $\theta \to ( \pi/2 )_-$ and to $+\infty$ as $\theta\to 0_+$. Following Ref.\ \cite{BrownEricG.2015Sasc}, we take $\theta(t)$ to be
\begin{align}
    \theta(t) &= \tan^{-1} \bigg( \frac{1 + e^{-\lambda t}}{\lambda} \bigg) 
    \vt 
\end{align}
as discussed in the main text. Since $0 < \theta(t) < \pi/2$, the mirror exists for all $t$, but it is never Dirichlet. It is perfectly transmissive in the asymptotic past, indicated by the limit $\theta(t) \to \pi/2$ as $t \to -\infty$, and its ``formation'' (i.e.\ its ability to reflect cavity modes) begins exponentially slowly. The end state of the mirror in the asymptotic future is not perfectly reflecting (Dirichlet), since $\theta(t) \to \mrm{cot}^{-1}(\lambda)$ as $t \to \infty$; however, it can be made arbitrarily close to Dirichlet by allowing $\lambda$ to be large.  To obtain the mode functions with frequency $k$, we make the ansatz,
\begin{align}
    U_k(u,v) &= \frac{1}{\sqrt{8\pi k}} \Big[ e^{-ikv} + E_k (u) \Big] 
    \vt 
\end{align}
where $v = t + x, u = t -x$ are lightcone coordinates and $E_k(u)$ is to be found. Note that the spatially odd solutions to the field equation Eq.\ (\ref{eq30}) do not feel the presence of the wall; only the spatially even solutions do. By spatial evenness, it suffices to consider these solutions in the half-space $x>0$, where those in the space $x<0$ follow by the replacement $(t,x) \to (t,-x)$. If we restrict ourselves, for simplicity, to the right-moving modes, then it can be shown that the constraints imposed by Eq.\ (\ref{eq33}) imply solutions of the form,
\begin{align}
    E_k(u) &= R_{k/\lambda}(\lambda u)
    \vt 
\end{align}
where
\begin{align}
    R_K (y) &= \begin{cases}
        e^{-iKy} & y \leq 0 
        \vt \\
        e^{-iKy} - \frac{2}{B(y)} \int_0^y B'(y') e^{-iKy'} \mrm{d} y' & 0 < y < 1 
        \vt \\
        - e^{-iKy} 
        \vt & y \geq 1
    \end{cases}
\end{align}
and $B(y)$ is a solution to $B'(y)/B(y) = \mrm{cot}(h(y))$ with the initial condition $B(0) = 1$, with $h(y)=\tan ^{-1}\left(1+e^{-y}\right)$. For this choice of $h(y)$, the right-moving modes during the switching are given by \cite{BrownEricG.2015Sasc}
\begin{align}
    \bar U_k(u) &= \frac{1}{\sqrt{8\pi k}}\frac{e^{-iku}}{1+ e^{\lambda u}} \bigg( 1 - \frac{\lambda+ik}{\lambda-ik} e^{\lambda u} \bigg).
    \label{eqc8}
\end{align}
Henceforth, we work with the cavity modes explicitly, which can be obtained from Eq.\ (\ref{eqc8}) by replacing the prefactor $( 8\pi k )^{-1/2}$ with $( 4\pi m )^{-1/2}$ where $m \in \mathds{Z}$ and restricting $k = \pi m/a$. Meanwhile, the standing wave solutions of the global cavity are simply $U_m(u) = ( 4\pi m)^{-1/2} e^{-i \pi m u/a}$. We can expand the subcavity mode operators $\h a_m$ in a basis of the global mode operators $( \h b_n, \h b_n^\dagger)$ via the Bogoliubov transformation,
\begin{align}
    \h a_m &= \sum_{n=1}^{+\infty} \Big( \alpha_{nm} \h b_n + \beta_{nm} \h b_n^\dagger \Big) 
    \vt 
\end{align}
where the Bogoliubov coefficients $\alpha_{nm}, \beta_{nm}$ are defined in the usual way via the Klein-Gordon inner product. In the vacuum of the unperturbed modes, the expectation value of the particle number of the new modes is,
\begin{align}
    \langle 0_B | \h a_m^\dagger \h a_m | 0_B \rangle 
    &= \sum_{n=1}^{+\infty} \big| \beta_{nm} \big|^2 
    \vt 
    \label{eq42}
\end{align}
It now remains to compute the Bogoliubov coefficients $\beta_{k\omega}$, which are given by
\begin{align}
    \beta_{nm} = \langle U_m^\star(u), \bar U_n (u) \rangle &= i \int_0^{a/2} \mrm{d} u \: \Big( U_m (u) \partial_u \bar U_n - \bar U_n \partial_u U_m  \Big) 
    \vt 
\end{align}
Upon integration, one obtains, 
\begin{align}
    \beta_{nm} &= \frac{1}{4\pi} \sqrt{\frac{n}{m}} \bigg( \frac{i e^{-ia\vartheta\lambda}}{\vartheta\lambda} \prescript{}{2}{F}_1 \Big( 1 - i \vartheta , 1 - i \vartheta , - e^{a\lambda/2} \Big) + \frac{1}{2\lambda} \Big( \psi^{(0)} ( - i \theta/2 ) - \psi^{(0)} ( - i \theta/2 + 1/2 ) \Big) \bigg)
    \nonumber \vt \\
    & - \frac{1}{4\pi} \sqrt{\frac{n}{m}} \frac{\lambda + i \pi  m/a}{\lambda - i \pi m/a} \bigg( \frac{i}{2\vartheta\lambda} \Big( 2 \mrm{Re} \Big[ e^{ia \vartheta\lambda} \prescript{}{2}{F}_1 \Big( 1, - i \theta , - i \theta + 1 , -e^{a\lambda/2} \Big) \Big] - \pi \vartheta\mrm{csch}(\pi\theta) - 1 \Big) 
    \nonumber \vt \\
    & + \frac{1}{2\theta\lambda} \Big( 2 \mrm{Re} \Big[ i e^{-ia\vartheta\lambda} \prescript{}{2}{F}_1 \Big( 1 , i \theta , i \theta + 1 , - e^{-a\lambda/2} \Big) \Big] - \vartheta \mrm{Re} \Big[ \psi^{(0)} ( - i \theta/2 ) - \psi^{(0)} ( - i \theta/2 + 1/2 ) \Big] \Big) \bigg)
    \label{eqc12}
\end{align}
where $\prescript{}{2}{F}_1(\alpha,\beta,\gamma,z)$ is the hypergeometric function, $\psi^{(0)}(z)$ is the PolyGamma function \cite{gradshteyn2014table}, and we defined $\vartheta = (m+n)\pi/(a\lambda)$. Substituting Eq.\ (\ref{eqc12}) into Eq.\ (\ref{eq42}) gives the average particle number $\langle \h N \rangle$ in Fig.\ \ref{fig:suppression}(b) of the main text. 

Taylor expanding the effective reflectivity for short times, we obtain for small $\lambda$, that its behaviour at short times scales as $O(-t \lambda^2)$. This indicates that a single order of magnitude reduction in $\lambda$, corresponds to a two order of magnitude reduction in the reflectivity rise time-scale. In this way, if we associate $\lambda = 20$ with a fast 10-100 femtosecond time-scale, $\lambda \sim 2 \times 10^{-3}$, corresponds to a micro-second scale of the switching time. This slower reflectivity switch produces particle content on the order of $5 \times 10^{-8}$, corresponding to a six order of magnitude reduction in the frequency shift. This indicates that a measurable tail of the frequency shift will survive that is larger than the linewidth of the control atom. Furthermore the toy model agrees with the behaviour of the reflectivity switch in the reflective branch of the superposition solved from the full model of microscopic interactions in Appendix~\ref{arrayresponse}, after associating $\lambda^2$ with the eigenvalues of the matrix of ODEs of the dynamics for the atomic amplitudes (in the sense of a power law suppression).

\subsection{Non-Perturbative Transition Calculations and Cavity Intensity}\label{nonperturb1}
If the control atom is driven off-resonance, there should be a substantially reduced probability for the control atom to flip to its reflective state. As expected for weak couplings, such that $g  \ll \nu \sim \omega_1$, the standard off-resonant suppression expected in the quantum optical Rabi model is observed, scaling as $g^2$:
\begin{equation}
    c_R = \frac{g^2}{\Omega^2 } \sin ^2\left(\Omega t / 2\right),
\end{equation}
where $\Omega = \left[(\delta + \delta_R)^2 + g^2 \right]^{1 / 2}$. This implies for the cavity extraction protocol, that when the cavity is driven on-resonance with the sub-cavity frequency $\omega_1$, that the average intensity observed in the cavity, will be suppressed by the non-zero probability for the atom array to be in the reflective state $|c_R|^2$. The intensity of light in the cavity close to the sub-cavity frequency is

 \begin{equation}
    I=\frac{I_{\max }|c_R|^2}{1+(2 \mathscr{F} / \pi)^2 \sin ^2\left(\pi \nu_P / \omega_1\right)},
 \end{equation}
 where $I_{\max }=\frac{I_0}{(1-r)^2}$, $\mathscr{F}$ is the finesse of the cavity, $r$ is the optical attenuation factor due to any imperfect reflectivity of the array and $\nu_P$ is the frequency of the pump. In this way, a substantial average power will only be observed in the cavity at the sub-cavity frequency, if the control atom was driven at the re-normalised frequency due to the presence of photons from entangled sub-regions of the vacuum of the global cavity.

\section{Comparison to the Dynamical Casimir Effect and the Lamb Shift} \label{DCEcomparison}
\subsection{Parametric Forms of the Dynamical Casimir Effect}\label{paracasimir}
In this section, we compare the particle creation phenomena considered in this work with parametric forms of the DCE, in which the effect can be described with a Bogoliubov expansion truncated to a single mode expansion. In the case for which the larger sub-cavity length is approximately the global cavity length, such that $\frac{r - L}{L} \ll 1$, the Bogoliubov relation between the modes can be truncated at lowest order in the expansion,
\begin{equation}
    \hat{a}_j \simeq \alpha_{jj} \hat{b}_j-\beta_{jj} \hat{b}_j^{\dagger},    
\end{equation}
In this case, the Hamiltonian of the sub-cavity transforms as:
\begin{equation}
    \hat{H} = \sum_j \omega_j \hat{a}_j^\dag \hat{a}_j \rightarrow \sum_j \omega_j (|\alpha_{jj}|^2 + |\beta_{jj}|^2) \hat{b}_j^\dag \hat{b}_j -  \omega_j\alpha_{jj}\beta_{jj}(\hat{b}_j^\dag\hat{b}_j^\dag + \hat{b}_j\hat{b}_j) + \omega_j|\beta_{jj}|^2,
\end{equation}
now to quadratic order in the Bogoliubov expansion we use that $|\alpha_{jj}|^2 - |\beta_{jj}|^2 \simeq 1$, in this perturbative regime, and we see that the Hamiltonian simplifies to the free-energy term of the global cavity, plus a squeezing term:
\begin{equation}
    \hat{H} \rightarrow \sum_j \omega_j \hat{a}_j^\dag \hat{a}_j \rightarrow \sum_j \omega_j  \hat{b}_j^\dag \hat{b}_j -  \omega_j\alpha_{jj}\beta_{jj}(\hat{b}_j^\dag\hat{b}_j^\dag + \hat{b}_j\hat{b}_j) + \omega_j|\beta_{jj}|^2.
\end{equation}
In this perturbative regime, we can see the time-dependent boundary condition produces parametric generation of pairs of photons in each cavity mode. If the cavity is driven with a drive frequency $\omega_D$, with the cavity initially in the ground state the probability to produce pairs of photons in the cavity is $P_j(\ket{n = 2}) = \omega_j^2\alpha_{jj}^2\beta_{jj}^2t^2\mathrm{sinc}^2((\omega_D - 2\omega_j)t)$. This corresponds to the resonant production of pairs of photons in the cavity, comparable to the parametric generation of cavity photons due to the motion of a mirror in blue-detuned regime of optomechanical set-ups \cite{RevModPhys.86.1391}. In this way, parametric forms of the DCE, can be interpreted as the parametric generation of photon pairs from a resonant drive as in optomechanical set-ups, not particle content due to entanglement between spatially distinct sub-systems, or particle creation from a fundamentally altered photonic mode structure from a time-dependent boundary condition. Indeed, all DCE experiments to date can be interpreted as the parametric coupling of unperturbed and time-independent cavity modes. In this way, our proposal allows probing fundamental physics neither tested nor discussed in the related albeit distinct context of the existing DCE experiments. Furthermore, we note that defining the detuning $\Delta = \omega_D - 2\omega_j$, we find for fixed interaction times, a quadratic suppression of the particle creation probability with the detuning, indicating the scaling of the suppression of particle creation for the adiabatic regime (slow switching).

More concretely, we model the Bogoliubov transformation with time-dependent Bogoliubov coefficients, modelling a time-dependent change in cavity length or equivalently refractive index or susceptibility of a medium: 
\begin{equation}
    \hat{H}=\sum_j \omega_j \hat{a}_j^{\dagger} \hat{a}_j,
\end{equation}
with time-dependent Bogoliubov transformation:
\begin{equation}\label{ajj}
    \hat{a}_j(t) \rightarrow \alpha_{j j}(t) \hat{b}_j-\beta_{j j}(t) \hat{b}_j^{\dagger},
\end{equation}
If we use this transformation in the Hamiltonian, the Hamiltonian transforms as 
\begin{equation}\label{transform111}
    \widetilde{H}(t)=S(t) \hat{H}_0(t) S^{\dagger}(t)-i  S(t) \dot{S}^\dagger (t),    
\end{equation}
such that $S(t)$ is a Bogoliubov transformation that implements the squeezing operation,  $\hat{a}_j= S(t) \hat{b}_j S^{\dagger}(t), \hat{a}_j^{\dagger}=S^{\dagger}(t) b_j^{\dagger} S(t)$:

\begin{equation}\label{SSdot}
    \hat{S}(t) \dot{\hat{S}}^{\dag}(t) = \frac{\dot{\zeta}^{\star}(t)}{2} \hat{b}^{\dagger 2}-\frac{\dot{\zeta}(t)}{2} \hat{b}^2,
\end{equation}
where $\alpha = \cosh (|\zeta(t)|) ,
\beta = -\sinh (|\zeta(t)|)$. Identifying the Bogoliubov coefficients with squeezing parameters, we can write the transformed Hamiltonian in Eq.~\eqref{transform111} as:
\begin{equation}
   \widetilde{H}(t)= \omega(\left|\alpha_{j j}\right|^2+\left|\beta_{j j}\right|^2)\left(b_j^{\dagger} b_j + \frac{1}{2}\right)+\hbar C(t) b_j^{\dagger 2}+\hbar C^*(t) b_j^2,
\end{equation}
where the combination of the additional term from Eq.~\eqref{SSdot}, as well as the transformation from Eq.~\eqref{ajj} give squeezing co-efficient
\begin{equation}
    C(t) = -\omega_j \alpha_{jj}(t) \beta_{jj}(t) - i\dot{\zeta}^{\star}(t).
\end{equation}
In the slow-switching regime, the first term here $\omega_j \alpha_{jj} \beta_{jj}$ corresponds to corrections to the free Hamiltonian and therefore the new vacuum state due to the time-dependent boundary conditions (the mismatch with the initial vacuum state which previously resulted in particle content), while the second term will result in non-adiabatic corrections, and is clearly a factor of $\frac{\gamma}{\omega_j}$ (where $\gamma$ is the slow rate of change of the time-dependent coefficients) smaller in terms of order of magnitude once the finite switching rate is take into account. In this way, the quantum state under the drive Hamiltonian evolves as (according to the adiabatic theorem, to lowest order in the unitary evolution),
\begin{equation}
    \ket{\psi(t)} \simeq  | 0_{a_j} \rangle \ket{T} + (gt) | 0_{\tilde b_j} \rangle | R \rangle 
\end{equation}
where $\hat{a}_j\ket{0_{a_j}} = 0$, and $| 0_{\tilde b_j} \rangle$ is the vacuum state with respect to the free Hamiltonian and adibatically shifting squeezing terms, up to the non-adiabatic corrections defined above. Importantly, such non-adiabatic corrections quantified as above produce the signature frequency shift by shifting out of this vacuum state by the factor $\frac{\gamma}{\omega_j}$.

\subsection{Comparison to the Lamb Shift}
Computing the energy shift of the excited state of the control atom due to emission or absorption of a virtual photon into the vacuum (as in the Lamb shift) produces a quantitatively distinct frequency shift. Most notably, the leading contribution to the Lamb shift occurs at second order in perturbation theory, whereas the shift we consider here is non-vanishing at leading order. Furthermore, the Lamb shift applies to energy corrections for processes where the input and output states are identical, whereas our analysis involves transitions between different Fock states of the photonic field. Higher order terms in the Dyson expansion also involve Lamb-shift-type energy corrections, that depend on the different energy distributions of the two sub-cavity regions. This can be seen if we take the second order term in the unitary evolution operator, which involve frequency shifts due to transitions between the same energy states of the control:

\begin{equation}
\begin{aligned}
   \hat{U}_I^{\mathrm{Lamb}} = & -\frac{g^2}{2} \int_0^t \mathrm{~d} t^{\prime} \int_0^{t^{\prime \prime}} \mathrm{d} t^{\prime \prime}\left(e^{i \hat{H}_T t^{\prime}}|T\rangle\langle R| e^{-i \hat{H}_R t^{\prime}} e^{i \hat{H}_T t^{\prime \prime}}|T\rangle\langle R| e^{-i \hat{H}_R t^{\prime \prime}} e^{i\left(\omega_D-\nu\right)\left(t^{\prime}+t^{\prime \prime}\right)}\right. \\
    & \quad+e^{i \hat{H}_T t^{\prime}}|T\rangle\langle R| e^{-i \hat{H}_R   t^{\prime}} e^{i \hat{H}_R t^{\prime \prime}}|R\rangle\langle T| e^{-i \hat{H}_T t^{\prime \prime}} e^{i\left(\omega_D-\nu\right)\left(t^{\prime}-t^{\prime \prime}\right)} 
    \vphantom{\int_0^{t'}}
    \\
    & \quad+e^{i \hat{H}_R t^{\prime}}|R\rangle\langle T| e^{-i \hat{H}_T t^{\prime}} e^{i \hat{H}_T t^{\prime \prime}}|T\rangle\langle R| e^{-i \hat{H}_R t^{\prime \prime}} e^{-i\left(\omega_D-\nu\right)\left(t^{\prime}-t^{\prime \prime}\right)}
    \vphantom{\int_0^{t'}}
    \\
    & \left.\quad+e^{i \hat{H}_R t^{\prime}}|R\rangle\langle T| e^{-i \hat{H}_T t^{\prime}} e^{i \hat{H}_R t^{\prime \prime}}|R\rangle\langle T| e^{i \hat{H}_T t^{\prime \prime}} e^{-i\left(\omega_D-\nu\right)\left(t^{\prime}+t^{\prime \prime}\right)}\right)+O\left(g^3\right),
    \vphantom{\int_0^{t'}}
\end{aligned}
\end{equation}
and we can see that this produces an energy shift that depends on the energy of the reflective and transmissive branches respectively, for transitions from the $\ket{T}$ state to the $\ket{T}$ state, as well as the $\ket{R}$ state to the $\ket{R}$ state. We leave a detailed analysis of the dynamics of such Lamb-shift type corrections to future work.

\subsection{Evidence for Vacuum Entanglement and Outlook}

While the vacuum particle content discussed here, either in the fast or slow switching regime is fundamentally a signature of genuine vacuum particle creation, in the fast-switching regime it can also be considered to be a signature of the entanglement structure of the vacuum. If the global cavity is initially prepared in a pure state—which can be set up and verified experimentally before driving the control—then the Schmidt criterion applies: Sub-systems whose matrices are not rank-1 are necessarily entangled. Measuring the rank in this way would require tomography of the quantum state of the sub-regions of the global cavity, which extracting from solely a measurement of the frequency shift of the control is an interesting problem beyond the scope of this work. However, we emphasize that sub-region particle content remains a testable prediction of the standard entanglement structure of the electromagnetic vacuum. Given projective measurements on the control atom, entangled photons across the two sub-cavities could also be measured after projection onto the reflective branch of the superposition using a standard entanglement witness for the entangled pairs of photons produced across the two sub-cavities. In this way, the experimental implementation of our proposal in the fast-switching regime would not only mark the first observation of the local particle content of the vacuum, a seminal yet hitherto untested prediction of QFT, but also serve as the first experimental signature of the entanglement structure of the vacuum. 

Our proposal further explores a novel regime of light–matter interaction by implementing dynamics of superpositions of macroscopically distinct photonic QED vacua, motivating the experimental realisation and exploration of fundamental phenomena in QFT. This platform also enables the observation of Rabi oscillations affected by the entangled vacuum's local particle content, offering a novel quantum-enhanced probe of the frame-dependent particle content central to QFT in curved spacetime \cite{Birrell_Davies_1982}, and it is an open question as to whether it could enable the observation of quantum reference frame \cite{GiacominiFlaminia2019Qmat} dependent particle content. In particular, it allows controlled access to the photon content perceived by observers in a superposition of trajectories \cite{PhysRevD.102.085013,PhysRevD.102.045002}. In general, the here-proposed platform enables a novel type of vacuum particle creation, in which the process is coherently superposed with the absence of particle creation. Our proposal extends the quantum-optical interaction of light and matter and state-dependent boundary conditions \cite{PhysRevA.56.3187,PhysRevA.72.022320,PhysRevLett.127.043603,Sinha_2025} (involving entanglement of a quantum control with an optical degree of freedom in a cavity QED setting) into the unexplored domain in which the quantum-control determines macroscopically distinct photonic vacuum structures, paving the way towards a new class of near-term experiments that bridge quantum optics, relativistic quantum information, quantum field theory and quantum sensing.

\bibliography{biblio} 
\section*{Acknowledgements} 
We thank the participants of the Swedish workshop on analogue gravity, as well as Navdeep Arya, Simon Hollerith, Robert Mann, Eduardo Martin-Martinez, Rick Perche, Igor Pikovski, and Johannes Zeiher for discussions. This material is based upon work supported by the Knut and Alice Wallenberg foundation through a Wallenberg Academy Fellowship No. 2021.0119, and the General Sir John Monash Foundation. J.F.\ acknowledges funding provided by the Natural Sciences and Engineering Research Council of Canada through a Banting Postdoctoral Fellowship. SQ is funded in part by the Wallenberg Initiative on Networks and Quantum Information (WINQ) and in part by the Marie Skłodowska--Curie Action IF programme \textit{Nonlinear optomechanics for verification, utility, and sensing} (NOVUS) -- Grant-Number 101027183. Nordita is supported in part by NordForsk. GT Thanks the Mainz Institute for theoretical physics and the Benasque quantum science centre for hospitality during the development of this work. 

\end{document}